\documentclass[preprint]{aastex}

\shorttitle{Mass and Semi major Axis Distribution of Extra Solar
Planets I: Three Planetary Populations} 

\shortauthors{Ida and Lin}

\begin{document}

\title{Towards a Deterministic Model of Planetary Formation I:
a Desert in the Mass and Semi Major
Axis Distributions of Extra Solar Planets}

\author{S. Ida}
\affil{Tokyo Institute of Technology,
Ookayama, Meguro-ku, Tokyo 152-8551, Japan}
\affil{UCO/Lick Observatory, University of California, 
Santa Cruz, CA 95064}
\email{ida@geo.titech.ac.jp}

\and 

\author{D. N. C. Lin}
\affil{UCO/Lick Observatory, University of California, 
Santa Cruz, CA 95064}
\email{lin@ucolick.org}

\begin{abstract}
In an attempt to develop a deterministic theory for planet formation,
we examine the accretion of cores of giant planets from planetesimals,
gas accretion onto the cores, and their orbital migration.  We adopt a
working model for nascent protostellar disks with a wide variety of
surface density distributions in order to explore the range of
diversity among extra solar planetary systems.  We evaluate the cores'
mass-growth rate $\dot{M}_{\rm c}$ through runaway planetesimal
accretion and oligarchic growth.  The accretion rate of cores is
estimated with a two-body approximation.  In the inner regions of
disks, the cores' eccentricity is effectively damped by the
ambient disk gas and their early growth is stalled by ``isolation''.
In the outer regions, the cores' growth rate is much
slower.  If some cores can acquire more mass than a critical value
of several Earth masses during the persistence of the
disk gas, they would be able to rapidly accrete gas and evolve into
gas giant planets.  The gas accretion
process is initially regulated by the Kelvin-Helmholtz contraction of
the planets' gas envelope.  Based on the assumption that the
exponential decay of the disk-gas mass occurs on the time scales $\sim
10^{6}-10^{7}$ years and that the disk mass distribution is comparable
to those inferred from the observations of circumstellar disks of T
Tauri stars, we carry out simulations to predict the distributions of
masses and semi major axes of extra solar planets. In disks
as massive as the minimum-mass
disk for the Solar system, gas giants can form only slightly outside
the ``ice boundary'' at a few AU.  But, cores can rapidly grow above
the critical mass interior to the ice boundary in protostellar disks
with 5 times more heavy elements than those of the minimum-mass disk.
Thereafter, these massive cores accrete gas prior to its depletion and
evolve into gas giants.  Unimpeded dynamical accretion of gas is a
runaway process which is terminated when the residual gas is depleted
either globally or locally in the form of a gap in the vicinity of
their orbits.  Since planets' masses grow rapidly from $10 M_{\oplus}$
to $100 M_{\oplus}$, the gas giant planets rarely form with asymptotic
masses in this intermediate range.  Our model predicts a paucity of
extra solar planets with mass in the range 10-$100 M_{\oplus}$ and
semi major axis less than 3AU.  We refer to this deficit as a ``planet
desert''.  We also examine the dynamical evolution of protoplanets by
considering the effect of orbital migration of giant planets due to
their tidal interactions with the gas disks, after they have opened up
gaps in the disks.  The effect of migration is to sharpen 
the boundaries and to enhance
the contrast of the planet desert. It also clarifies the separation
between the three populations of rocky, gas-giant, and ice-giant
planets.  Based on our results, we suggest that the planets' mass
versus semi major axes diagram can provide strong constraints on the
dominant formation processes of planets analogous to the implications
of the color-magnitude diagram on the paths of stellar evolution.
Finally, we show that the mass and semi major axis distributions
generated in our simulations for the gas giants are consistent with
those of the known extra solar planets.  Future observations can
determine the existence and the boundaries of the planet desert in
this diagram, which can be used to extrapolate the ubiquity of rocky
planets around nearby stars.

\end{abstract}


\section{Introduction}
\label{sec:introduction}

More than 100 extra solar planets have been discovered so far.  This
sample is sufficiently large that some clues and constraints on the
formation of extra solar planets may be inferred from their kinematic
properties.  Figure~\ref{fig:obs} shows the distributions of semi
major axes ($a$) and masses ($M_{\rm p}$) of the known extra solar
planets as of July 2003 (http://www.obspm.fr/encycl/encycl.html).
Various radial-velocity surveys may have exhaustively found most of
the planets in the solar neighborhood with radial velocity amplitude
$v_r \ga 10$ ms$^{-1}$ and relatively small semi major axes, $a \la 4$-5AU,
which correspond to orbital periods $\la 10$ years.  The mass
distribution in these regions appears to follow a weak power-law
function in $M_{\rm p}$ with an abrupt upper cut-off at around 10
$M_{\rm J}$, \citep{Jorissen01, Tabachnik02, Zucker02} where $M_{\rm
J}$ is a Jupiter mass and is $\simeq 320M_{\oplus}$ ($M_{\oplus}$ is
an Earth mass).

But, Figure~\ref{fig:obs} also shows a correlation between the $M_{\rm
p}$ and $a$ distributions. \citet{Zucker02}, \citet{Patzold02} and
\citet{Udry03}
discussed a deficit of massive ($M_{\rm p}\ga 5M_{\rm J}$) and short
period ($a\la0.2$AU) planets.  Here we focus on another domain of
paucity in the $M_{\rm p}-a$ distribution: the regions $a \ga 0.2$AU
and $M_{\rm p} \la M_{\rm J}$ (Figure~\ref{fig:obs}).  This paucity is
not due to the observational selection effects of precision and
legacy baseline since planets in this domain have $v_r \ga 10$ ms$^{-1}$ and
orbital periods $P$ less than a few years \citep{Udry03}.  
In this paper, we suggest
that this paucity reflects the condition for and the process of
planetary formation.

We interpret the observational data on Jupiter-mass extra solar
planets with the conventional core accretion model of gas giant
planets (Wuchterl et al. 2000 and references therein).  In the
standard model, terrestrial planets and solid cores of gas giants are
formed through the coagulation of planetesimals with initial sizes
$\sim$ 1-10 km, e.g., \citep{Safronov69,Wetherill80,Hayashi85}.  
Cores which are heavier than a few lunar masses
attract the disk gas and form hydrostatic atmospheres. But, if a solid
core reaches a critical mass of 
$\ga$ several $M_{\oplus}$ (\S 3.1), the pressure
gradient in planet's gaseous atmosphere can no longer support against
its gravity and the planet's atmosphere collapses onto its core
\citep{Mizuno80, BP86, P96, Ikoma00}.  Subsequent disk gas accretion
onto the solid core forms a gas giant planet.

The cores' asymptotic masses and their accretion time scales are
determined by surface density of the planetesimal swarm and orbital
radius \citep{KI02}.  In general, the formation of the gas giants is
favored slightly exterior to the ``ice boundary'' at a few AU \citep{KI02}, 
where the surface density of dust particles increases by a factor three to
four as ice condenses.  This enhancement of the dust surface density
facilitates the formation of large cores.  Well beyond the ``ice
boundary'' ($\ga 10$AU), the cores' growth rate is so small that the
residual gas in the disk tends to be depleted before it can be
accreted.  In very massive protoplanetary disks, however, cores can
reach the critical mass interior to the ice boundary
\citep{LI97, KI02}.  Since the core
accretion is relatively fast in the inner regions of the disk, gas
accretion onto cores can start well before the gas is depleted.
Unimpeded dynamical accretion of gas is a runaway process which is
terminated when the residual gas is depleted either globally or
locally in the form of a gap in the vicinity of the gas giants'
orbits.  But, prior to a self-limiting growth stage, the planets grow
so rapidly from $10 M_{\oplus}$ to more than $100M_{\oplus}$, that
they rarely emerge with an asymptotic mass in the intermediate range
($10-100 M_{\oplus}$).  We will refer to such a planet-deficit domain
in $M_{\rm p}$ versus $a$ distribution as a ``planet desert''.

After their formation, gas giant planets with sufficiently large
masses to open up gaps also undergo type II orbital migration which is
well coupled to the viscous evolution of their nascent
disks \citep{LP85}.  If these planets' tidal and magnetospheric
interaction with their host stars can halt their inward migration,
there would be an accumulation of ``short-period planets''
(equivalently ``close-in planets'', or ``hot Jupiters''), near the
disk inner edge ($\la 0.1$AU).  Dozens of short-period planets have been
discovered \citep{Marcy00} and their masses appear to be relatively
small \citep{Udry03}.  These planets have either formed with relatively low
masses (see \S 4) or lost some of their initial mass through Roche
lobe overflow \citep{Trilling98, Gu03}. The latter process is only
effective in the regions close to the stellar surface.  
Massive short-period planets
may also be more vulnerable to post formation orbital decay due to
their enhanced tidal interaction with their host stars
\citep{Zucker02, Patzold02}.

Gas giant planets may attain $a$ in the range of at $0.2-3$AU, only if
the disk gas is timely depleted during their migration \citep{Trilling02, 
Armitage02}.  Type II migration proceeds at a faster pace for planets
with relatively lower mass and smaller orbital radius.  As shown in
\S5, the intermediate-mass planets can open up gaps in inner
regions ($a \la 1$AU) and easily migrate to the proximity of the host
stars.  Larger planets tend to migrate less rapidly than the
intermediate-mass planets.

Intermediate-mass planets with insufficient masses to open gaps also
interact tidally with the disk gas through their Lindblad and
corotation resonances \citep{GT80}.  In principle, any imbalance in the
torque can lead to ``type I'' migration \citep{Ward86}. The
effect of the corotation resonances \citep{Tanaka02}  and secondary
instabilities in the vicinity of the coorbital region \citep{Balmforth01}
may be able to quench the magnitude and sign of the net torque.
Observationally, there is no evidence that type I migration has
brought in any water-filled terrestrial planets into the inner Solar
system.

Well outside the ice boundary ($\ga 3$AU), the intermediate-mass
planets can slowly accrete nearby planetesimals.  Prior to the
depletion of the gas, however, intermediate-mass planets are unable to
open up gaps in outer regions.  In the absence of type I migration,
they cannot migrate to the inner region from larger orbital radii.
Cores in the outer regions of the disk can attain masses in excess of
$1 M_{\oplus}$ after the gas depletion (\S 2).  The mass
distribution for long-period planets are expected to be continuous.
But, with the residual gas depleted, the intermediate-mass planets
which formed lately cannot migrate to fill the planet desert. 

Based on these considerations, we suggest that the intermediate-mass
planets may be rare and their deficit gives rise to a ``planet desert''
at $0.2-3$ AU.  In \S2, we present the model of core accretion and
predict cores' masses as functions of their nascent disks' mass and
semi major axis.  In \S3, we include the effects of gas accretion onto
the cores.  Assuming an appropriate distribution of disk masses that
may be comparable to the observationally inferred mass distribution, we
carry out, in \S 4, Monte Carlo calculations to simulate the mass and
semi major axis distributions of extra solar planets.  The predicted
distributions clearly show a paucity of planets with masses $\sim 10-100
M_{\oplus}$ at $a \la 3$AU.  \S 5 discusses the effects of radial
migration of giant planets due to their tidal interaction with a gas
disk. We show that type II migration sharpens the boundaries and
enhances the contrast of the planet desert.  In \S6, we
summarize our results.

\section{Core Accretion}
\label{sec:core}

In this section, we briefly recapitulate the current theories for
solid core accretion.  Our main objective is to determine the cores'
growth rate and asymptotic masses.

\subsection{Core accretion model: the feeding zone}

In an early stage of planetesimal accretion, the largest planetesimals
have the shortest mass-doubling time scale. This process is commonly
referred to as the 'runaway growth', e.g., 
\citep{Greenberg78, WS89, ALP93, KI96}.  
In the limit that their velocity
dispersion is relatively small, field planetesimals are readily
captured by nearby runaway bodies (cores).  However, the differential
Keplerian speed of the planetesimals prevents them from reaching the
distant runaway cores until they have acquired sufficient orbital
eccentricity.  The region within which that the runaway cores can
directly accrete planetesimals is commonly referred to as the `feeding
zone'.  When the runaway cores capture all the planetesimals in their
feeding zones, their growth stops as they acquire an 'isolation
mass'.  \citet{Lissauer87} evaluated the isolation mass as a function
of the initial surface density of the solid components in protostellar
disks.  Assuming a negligible velocity dispersion for the
planetesimals, the full width of the feeding zone is given by 
$\Delta a_{\rm c} \simeq 2 \sqrt{12}r_{\rm H} \simeq 7r_{\rm H}, $ where
\begin{equation}
r_{\rm H} \equiv \left( \frac{M_{\rm c}}{3M_*} \right)^{1/3}a,
\label{eq:rH}
\end{equation}
is the Hill radius (Roche lobe radius) for a solid core with a mass
$M_{\rm c}$ around a host star with a mass $M_*$.

However, when a core becomes sufficiently massive to excite the
eccentricity and to pump up velocity dispersion of the neighboring
planetesimals, its feeding zone expands.  The increases in the
relative speed between the planetesimals and the cores also reduce the
effective cross section of the cores such that their growth slows down
\citep{IM93, ALP93, Rafikov03}.  The relatively massive cores are more
effective in stirring up the velocity dispersion of their neighbors so
that the reduction of their effective cross section is larger.
Consequently, the growth time scale becomes an increasing function of
the cores' mass.  Such a self-regulated growth results in the
formation of a population of comparable-mass cores which are embedded
in a swarm of small and relatively eccentric planetesimals
\citep{KI98, KI00, Thommes02}.  In this system, the orbital spacings
of the cores (equivalently, full width of their feeding zone, $\Delta
a_{\rm c}$) are nearly equal to one another and are as large as $\sim 10 r_{\rm H}$.
This maximization of spacing is the combined consequence of the
distant perturbations between the cores and the effect of dynamical
friction from the small planetesimals on the cores \citep{KI95}.
\citet{KI98} called this coagulation process, the 'oligarchic growth'.
Because the width of the feeding zone is
\begin{equation}
\Delta a_{\rm c} \sim 10 r_{\rm H}
\end{equation}
during the oligarchic growth stage, the isolation masses are 
slightly larger
than those estimated by \citet{Lissauer87}.

The concept of an isolation mass induced by the depletion of the
feeding zone is based on the assumption that the orbits of the 
cores do not evolve with time. In reality, the cores undergo radial
diffusion due to their scattering of the field planetesimals, e.g.,
\citep{Murray98,Ida00} and their tidal interaction with the
gas disk (type I migration), e.g., \citep{GT80,Ward86}.
Under the action of gas drag, the small planetesimals and dust
particles may also undergo migration, e.g., \citep{Adachi76}.  These
effects can lead to an expansion of the feeding zones.  Thus, in a
prescription for the asymptotic mass of cores' growth, it is
worthwhile to consider the possibility that the cores' feeding zone
may be extended throughout the disk.

\subsection{The cores' growth rate}
The growth rate of the cores is determined by the velocity dispersion
($\sigma$) or equivalently the eccentricity distribution of the
planetesimals and cores. Eccentricity damping due to their tidal
interaction with the disk gas \citep{Artymowicz93, Ward93} inhibits the
cores from attaining any significant growth in their velocity
dispersion.  In contrast, the eccentricity of the low-mass
planetesimals is less effectively damped.  Instead, it is excited by
cores through recoil of dynamical friction.  Through a series of 3D numerical
simulations, \citet{KI02} showed the planetesimals' eccentricity is
larger than the ratio of the cores' $r_{\rm H}$
and their semi major axes so that the cores' growth time scale is well
described by a simple two-body approximation.
 
In the modest to high-$\sigma$ limit, the cores' accretion rate at an
orbital radius $a$ is \citep{Safronov69}
\begin{equation}
\begin{array}{ll}
{\displaystyle
\dot{M}_{\rm c} }
&
{\displaystyle
\sim \pi R_{\rm c}^2 \rho_{\rm d} 
  \left( \frac{2GM_{\rm c}}{R_{\rm c}\sigma^2} \right) \sigma
\sim 2\pi R_{\rm c}^2 \Sigma_{\rm d} \Omega_{\rm K} 
  \left( \frac{GM_{\rm c}}{R_{\rm c}\sigma^2} \right) } \\
  & {\displaystyle
\sim 2\pi \left( \frac{R_{\rm c}}{a} \right)
\left( \frac{M_{\rm c}}{M_*} \right)
\left( \frac{a \Omega_{\rm K}}{\sigma} \right)^2
\Sigma_{\rm d} a^2 \Omega_{\rm K},
}
\end{array}
\label{eq:mdot}
\end{equation}
where $\Omega_{\rm K} = (GM_*/a^3)^{1/2}$ is a Kepler frequency,
$\rho_{\rm d}$ and $\Sigma_{\rm d}$ are the spatial and surface
density of solid components ($\rho_{\rm d} \sim \Sigma_{\rm d}
\Omega_{\rm K}/\sigma$), and $R_{\rm c}$ is a physical radius of the
core. In order for the planetesimals within $\Delta a_{\rm c}$ to
reach the cores,
\begin{equation}
\frac{\sigma}{\Omega_{\rm K}} \sim \Delta a_{\rm c} 
\sim 10 \left( \frac{M_{\rm c}}{3M_*} \right)^{1/3} a
\end{equation}
so that the solid cores' accretion time scale is
\begin{equation}
\tau_{\rm c,acc} = \frac{M_{\rm c}}{\dot{M}_{\rm c}} 
\sim \left( \frac{a}{R_{\rm c}} \right)
\left( \frac{M_{\rm c}}{\Sigma_{\rm d} a^2} \right)
\left( \frac{M_{\rm c}}{M_*} \right)^{-1/3} T_{\rm K},
\end{equation}
where $T_{\rm K} = 2\pi / \Omega_{\rm K}$ 
$[= (M_*/M_{\odot})^{-1/2}(a/1{\rm AU})^{3/2}$ years] is the Kepler period. 
By taking into account the effect of gas drag on the planetesimals'
velocity dispersion, \citet{KI02} derived a more detailed expression
(see their eqs.15 and 16),
\begin{equation}
\begin{array}{ll}
{\displaystyle
\tau_{\rm c,acc} \simeq} &
{\displaystyle
1.2 \times 10^5 
\left( \frac{\Sigma_{\rm d}}{10 \mbox{gcm}^{-2}} \right)^{-1}
\left( \frac{a}{1{\rm AU}} \right)^{1/2}
\left(\frac{M_{\rm c}}{M_{\oplus}} \right)^{1/3} 
\left( \frac{M_*}{M_{\odot}} \right)^{-1/6}} \\
  &
{\displaystyle
\times \left[ 
\left( \frac{\Sigma_{\rm g}}{2.4 \times 10^3 \mbox{gcm}^{-2}} \right)^{-1/5}
\left( \frac{a}{1{\rm AU}} \right)^{1/20}
\left( \frac{m}{10^{18}{\rm g}} \right)^{1/15}
\right]^{2}
{\rm years},
}
\end{array}
\label{eq:T_core_grow}
\end{equation}
where $\Sigma_{\rm g}$ is the surface density of the gas in the disk,
$m$ is the mass of typical planetesimals, and $M_{\odot}$ is the Sun's
mass.  Since $\tau_{\rm c,acc} \equiv M_{\rm c}/\dot{M}_{\rm c}$, we
find from eq.~(\ref{eq:T_core_grow})
\begin{equation}
\begin{array}{ll}
M_{\rm c}(t) \simeq &
{\displaystyle
\left( \frac{t}{3.5 \times 10^5{\rm years}} \right)^3 
\left( \frac{\Sigma_{\rm d}}{10 \mbox{gcm}^{-2}} \right)^{3}
\left( \frac{\Sigma_{\rm g}}{2.4 \times 10^3 \mbox{gcm}^{-2}} \right)^{6/5}
\left( \frac{m}{10^{18}{\rm g}} \right)^{-2/5}} \\
  & 
{\displaystyle
\times \left( \frac{a}{1{\rm AU}} \right)^{-9/5} 
\left( \frac{M_*}{M_{\odot}} \right)^{1/2}
M_{\oplus}},
\label{eq:m_grow}
\end{array}
\end{equation}
where we neglect variations of $\Sigma_{\rm d}$ and $\Sigma_{\rm
g}$ with time.  This approximation is adequately satisfied in the
inner regions of the disk where $\tau_{\rm c, acc}$ is small compared
with the gas depletion time scale.  

In the outer regions of the disk, the rate of mass growth is reduced
from the above expression during the depletion of the disk gas.  In
eq.~(\ref{eq:gas_decay}), we introduce a prescription for the global
depletion of the gas.  For this prescription of $\Sigma_{\rm g}$,
the evolution of the core mass becomes
\begin{equation}
\begin{array}{ll}
M_{\rm c}(t) \simeq &
{\displaystyle
\left( \frac{\tau_{\rm disk}}{1.4 \times 10^5{\rm years}} \right)^3 
\left( \frac{\Sigma_{\rm d}}{10 \mbox{gcm}^{-2}} \right)^{3}
\left( \frac{\Sigma_{\rm g, 0}}
{2.4 \times 10^3 \mbox{gcm}^{-2}} \right)^{6/5}
\left( \frac{m}{10^{18}{\rm g}} \right)^{-2/5}} \\
  & 
{\displaystyle
\times \left( \frac{a}{1{\rm AU}} \right)^{-9/5} 
\left( \frac{M_*}{M_{\odot}} \right)^{-1/2} \left( 1 - \exp 
\left( {- 2 t \over 5 \tau_{\rm disk}} \right)^3 \right) M_{\oplus}}
\label{eq:m_lategrow}
\end{array}
\end{equation}
where $\tau_{\rm disk}$ is the disk depletion time scale.  In the
limit of $t \ll \tau_{\rm disk}$, eq.~(\ref{eq:m_lategrow}) reduces to
eq.~(\ref{eq:m_grow}).

According to eq.~(\ref{eq:m_lategrow}), the cores attain an asymptotic
mass.  But for $t > \tau_{\rm disk}$, the assumptions for circular orbits of
the cores and damping of planetesimals' $\sigma$ by gas drag break down.  
When all the gas is depleted and all the heavy
elemental content is retained by the oligarchic cores, their $\sigma$
increases to become comparable to their characteristic surface speed 
$\sigma \sim V_{\rm surf} = (G M_{\rm c}/R_{\rm c})^{1/2}$ 
and their collisional cross
section reduces to their physical cross section.  In this limit, the
growth time scale for the final planetary embryos become
\begin{equation}
\tau_{\rm e, acc} 
\sim \left( { a \over R_{\rm e}} \right)^2 
\left( {M_{\rm e} \over \Sigma_{\rm d} a^2} \right)
\left({ T_{\rm K} \over 2 \pi^2} \right) 
\sim 4 \times 10^7 \left( {\Sigma_{\rm d} \over 
10 {\rm g cm}^{-2}} \right)^{-1}
\left( {M_{\rm e} \over M_\oplus} \right)^{1/3} 
\left( {\rho_{\rm e} \over 1 {\rm gcm}^{-3}} \right)^{2/3} 
\left( {{M_*} \over {M_{\oplus}}} \right)^{1/2} 
\left( a \over 1 {\rm AU} \right)^{3/2} {\rm years},
\label{eq:mdotembr}
\end{equation}
where $M_{\rm e}$, $R_{\rm e}$, and $\rho_{\rm e}$ are
a mass, radius, and an internal density of the embryo.
Here, we refer to solid planets after depletion of the gas as
"embryos" rather than "cores". 
The reduced collisional cross section significantly lengthens the growth time
scale for the embryos from that of the cores.
The corresponding mass for the embryos is
\begin{equation}
M_{\rm e}(t) \simeq M_{\rm e}(\tau_{\rm disk}) 
+ \left({t - \tau_{\rm disk} \over 6 \times 10^7 {\rm years}} \right)^3 
\left( {\Sigma_{\rm d} \over 10 {\rm g cm}^{-2} } \right)^3 
\left( {\rho_{\rm e} \over 1 {\rm gcm}^{-3}} \right)^{-2} 
\left({a \over 1 {\rm AU}} \right)^{-9/2} M_\oplus.
\label{eq:m_finalgrow}
\end{equation} 

\subsection{Disk structure models: the temperature and surface density 
distributions.}

The cores and embryos' mass, as expressed in both
eqs.~(\ref{eq:m_lategrow}) and (\ref{eq:m_finalgrow}), is a function of
the surface density of the heavy elements in the form of condensed
grains and planetesimals, $\Sigma_{\rm d}$.  The distribution of
$\Sigma_{\rm d}$ is poorly determined theoretically and
observationally.  In principle, the $\Sigma_{\rm d}$ distribution is
determined by the infall pattern of the heavy elements, the
condensation, growth, and sublimation rates of grains, and the
particles' interaction with ambient disk gas.  Here, we adopt an {\it
ad hoc} power-law distribution such that
\begin{equation}
\Sigma_{\rm d} = f_{\rm d} \times \eta_{\rm ice} 
                 \times 10 \left( \frac{a}{1{\rm AU}} \right)^{-3/2}
\mbox{ gcm}^{-2},
\label{eq:sigma_dust}
\end{equation}
based on a phenomenological minimum-mass solar nebula model
\citep{Hayashi81}.  The power-law dependence may be the consequence of
self-similar evolution.  We use this prescription as a fiducial model.
We plot $\Sigma_{\rm d}$ as a function of $a$ for $f_{\rm d} = 0.1,
1,$ and 10 in Fig.~\ref{fig:sig_dust_distr}.  Using these models, we
assess the formation properties of planetary systems as a function of
$\Sigma_{\rm d}$.  The step-function variable $\eta_{\rm ice}$ is
introduced to express the effect of ice condensation/sublimation
across the ice boundary ($a_{\rm ice} \sim$ a few AU; see 
eq.~[\ref{eq:a_ice}]).  For the inner
disk with $a < a_{\rm ice}$, $\eta_{\rm ice} = 1$ and for outer
regions with $a > a_{\rm ice}$, $\eta_{\rm ice} \sim 3-4$.  Following
the minimum-mass solar nebula model, we adopt $\eta_{\rm ice} = 4.2$
to represent the Solar abundance distribution \citep{Hayashi81}.

We also introduce a scaling parameter for the total disk mass, $f_{\rm
d}$, which is assumed to be a constant parameter throughout the disk.
In the case of the minimum-mass disk model for the Solar system
\citep{Hayashi81}, the scaling parameter $f_{\rm d} \simeq 0.7$.  This
value would be larger if the formation of the solid cores and their
retention into the Solar system planets were not highly efficient.
Radio observations of the continuum radiation in the mm-wave length
range measure the dust emission from the disks.  The observationally
inferred total mass of dust, $M_{\rm d}$, in the protostellar disks
around classical T Tauri stars ranges from $10^{-5} M_{\odot}$ to $3
\times 10^{-3}M_{\odot}$, e.g., \citep{BS96}.  The disk mass determinations
are somewhat uncertain due to the poorly known radiative properties of
the grains.  Nevertheless, their differential dust content provides a
reasonable evidence for a greater than an orders of magnitude
dispersion in $M_{\rm d}$.  In addition, the image of the disks is not
well resolved in many cases.  A rough magnitude of $f_{\rm d}$ can be
inferred from the the total mass of the disk under the assumption that
all disks have similar sizes (a few tens AU to a hundred AU).  The
observationally inferred mass of T Tauri disks corresponds to a range
of $f_{\rm d} \sim 0.1-30$.  We note that disks with $f_{\rm d}$
substantially larger than unity do commonly exist.

The presence of gas affects the velocity dispersion and therefore the
growth rate of the cores.  It is also important for the formation of
gaseous giant planets.  In principle, the magnitude of the gas surface
density $\Sigma_{\rm g}$ is determined by the gas infall rate and it
evolves as a consequence of the viscous stress in the disk, e.g.,
\citep{LP85}.  Similar to all astrophysical accretion disks, the
effective viscosity is generally assumed to be due to turbulence.
But, in regions of protostellar disks where planets emerge, the cause
of turbulence is highly uncertain \citep{PL95}.  For computational
simplicity, we also assume that $\Sigma_{\rm g}$ has a similar power
law dependence on $a$ as $\Sigma_{\rm d}$ such that
\begin{equation}
\Sigma_{\rm g} = f_{\rm g}
     \times 2.4 \times 10^3 \left( \frac{a}{1{\rm AU}} \right)^{-3/2}
\mbox{ gcm}^{-2}.
\label{eq:sigma_g}
\end{equation}
We vary the scaling parameter $f_{\rm g}$ to obtain different disk
models and to test the dependence of the planetary properties on the
mass of the gas disk.  This assumption implies that the relative ratio
of gas (volatiles) to dust (refractories composed mainly of heavy
elements)
\begin{equation}
\frac{\Sigma_{\rm g}}{\Sigma_{\rm d}} =  
\frac{f_{\rm g}}{f_{\rm d}} \frac{240}{\eta_{\rm ice}},
\end{equation}
is uniform throughout the disk
except across the discontinuity at the ice boundary.  While $f_{\rm
d}/f_{\rm g}=1$ would represent the Solar abundance, a different
choice of $f_{\rm d}/f_{\rm g}$ would represent the relative
enrichment/depletion of the heavy elements in the disks.

In the above equations, we introduce a separate mass scaling parameter
$f_{\rm g}$ for the gas in order to consider the possibility that the
solid components may evolve independently from the gas.  The basic
concept of the minimum-mass solar nebula model is based on the
assumption that all the heavy elements are retained while the original
disk gas was depleted. Theories suggest that gas diffuses under the
action of viscous stress, but dust and planetesimals migrate under the
influence of the gas drag effect such that they tend to evolve
independently from each other. Gas may also be preferentially depleted
through the photo evaporation e.g., \citep{Shu93, Johnstone03} and winds
while dusty grains may be preferentially retained by the
disk \citep{Shang00}.

At the moment, we can observe the dust component around most, and
molecular hydrogen around a few, young stars.  Among the classical T
Tauri stars, there is no apparent dependence of the mass of the dust
disks on the host stars' ages. For stars with ages greater than $\sim
10^{6}- 10^{7}$ years, the dust mass inferred from the mm-wave length
continuum is much reduced \citep{BS96, Wyatt03}.  The near-IR
signatures of disks also evolves on a similar time scale
\citep{HLL01}.  This decline has been interpreted as an evidence for
dust growth and planetesimal formation rather than the depletion of
heavy elements \citep{DAlessio01}. In this paper, we adopt
the conjecture that the surface density distribution of the dust plus
planetesimals does not change except in those regions where they have
been totally accreted by the cores.

Very little information is available on the evolution of $\Sigma_{\rm
g}$ and $f_{\rm g}$, but there is no indication of divergent
depletion pattern between molecular hydrogen and mm-size dust emission
\citep{Thi01}.  In addition, the presence of spectral evidences and UV
veiling for ongoing accretion are well correlated with the mm flux for
the classical T Tauri stars.  Both signatures of protostellar disks
also disappear together among weak line T Tauri stars \citep{Duvert00}.  
Based on these observations, we adopt the most
simple-minded assumption that gas is depleted on a similar time scale
even though the gas and dust depletion proceed through different
mechanisms.

In eq.~(\ref{eq:sigma_dust}) we introduce a parameter $\eta_{\rm ice}$
to take into account of the possibility of ice condensation.  In order
to determine the radius of the ice condensation, $a_{\rm ice}$, we need
to specify the gas temperature in the disk.  In most planet-forming
regions of interest, the gas density is sufficiently large to maintain
a thermal coupling between the gas and the dust particles.  In the
optically thin regions of the disk, the dust and the gas are heated to
an equilibrium temperature \citep{Hayashi81} such that
\begin{equation}
T = 280 \left( \frac{a}{1{\rm AU}} \right)^{-1/2}
  \left( \frac{L_*}{L_{\odot}} \right)^{1/4} {\rm K}
  \simeq 280 \left( \frac{a}{1{\rm AU}} \right)^{-1/2}
  \left( \frac{M_*}{M_{\odot}} \right) {\rm K}.
\label{eq:T_a}
\end{equation}
For solar-type stars, the stellar luminosity $L_*$ is roughly
proportional to $M_*^4$.  Since $a_{\rm ice}$ corresponds to the
radius where $T \simeq 170$K, we find that 
\begin{equation}
a_{\rm ice} = 2.7 \left( \frac{M_*}{M_{\odot}} \right)^{2} {\rm AU}.
\label{eq:a_ice}
\end{equation}
Hereafter, we adopt $a_{\rm ice} = 2.7$AU ($L_* = L_{\odot}$) until
\S 4.1.  In the Monte Carlo calculations in \S 4.2 and 5,
we consider variation of $a_{\rm ice}$.  Since scale
height of a gas disk is $h \simeq c_s/\Omega_{\rm K}$ where $c_s$ is
sound velocity, from eq.~(\ref{eq:T_a}), we find
\begin{equation}
\frac{h}{a} = 0.05 \left( \frac{a}{1{\rm AU}} \right)^{1/4}.
\label{eq:h_a}
\end{equation}

\subsection{The cores' asymptotic mass}

Based on the distributions of $\Sigma_{\rm d}$ given by
eq.~(\ref{eq:sigma_dust}), we find from eq.~(\ref{eq:m_grow}),
\begin{equation}
M_{\rm c}(t) \simeq 
\left( \frac{t}{3.5 \times 10^5{\rm years}} \right)^3 \eta_{\rm ice}^{3}
f_{\rm d}^{3} f_{\rm g}^{6/5}
\left( \frac{a}{1{\rm AU}} \right)^{-81/10} 
\left(\frac{M_*}{M_{\odot}} \right)^{1/2}
M_{\oplus}
\label{eq:m_grow0}
\end{equation}
by assuming $m=10^{18}$g.
Equivalently, the core accretion time scale is
(eq.~[\ref{eq:T_core_grow}])
\begin{equation}
      \tau_{\rm c,acc} \simeq 1.2 \times 10^5 
      \eta_{\rm ice}^{-1} f_{\rm d}^{-1} f_{\rm g}^{-2/5} 
      \left( \frac{a}{1{\rm AU}} \right)^{27/10}
      \left(\frac{M_{\rm c}}{M_{\oplus}} \right)^{1/3} 
      \left(\frac{M_*}{M_{\odot}} \right)^{-1/6} \mbox{ years}.
      \label{eq:pl_acc_rate}
\end{equation}
Note that the actual time scale to reach $M_{\rm c}$ is $3\tau_{\rm
c,acc}$ but not $\tau_{\rm c,acc}$.  The numerical factor of 3 comes
from the dependence that $\tau_{\rm c,acc} \propto M_{\rm c}^{1/3}$.

The unimpeded core growth which is described by eq.(\ref{eq:m_grow0}) 
can not be sustained indefinitely.
In the limit that damping effect due to disk-planet 
interaction \citep{Artymowicz93, Ward93} regulates a small velocity
dispersion for the cores, their growth terminates when they consume
all the planetesimals in their feeding zones.  If the radial
diffusion/migration of the cores is neglected, the isolation mass
\citep{Lissauer87, KI98, KI00} 
\begin{equation}
M_{\rm c,iso} = 2 \pi a \Delta a_{\rm c} \Sigma_{\rm d} 
\end{equation}
represents the asymptotic mass of the cores prior to the depletion of
the disk gas.  If we adopt $\Delta a_{\rm c} = 10 r_{\rm H}$ (where
$M_{\rm c}$ in $r_{\rm H}$ is replaced by $M_{\rm c,iso}$), the
isolation mass would be
\begin{equation}
M_{\rm c,iso} \simeq
0.16 \left(\frac{\Sigma_{\rm d}}{10 \mbox{gcm}^{-2}}\right)^{3/2}
\left(\frac{a}{1\mbox{AU}}\right)^{3}
\left(\frac{\Delta a_{\rm c}}{10r_{\rm H}} \right)^{3/2} 
\left( \frac{M_*}{M_{\odot}} \right)^{-1/2} M_{\oplus}.
\label{eq:m_iso}
\end{equation}

Upon acquiring $M_{\rm c,iso}$, the cores consume all the nearby
planetesimals.  These cores cannot cross each other's orbits,
until the gaseous-disk mass has decreased to $\la 10^{-5}M_\odot$, 
equivalently $f_{\rm g} \la 10^{-3}$, and the damping effect 
due to disk-planet interaction becomes weak enough 
\citep{Iwasaki02, Kominami02}. Even in a gas free environment, a swarm
of planetesimals has a tendency to evolve into a quasi-static oligarchic state.
\citet{KI02} performed three-dimensional N-body (with N=10,000)
simulations of planetesimal accretion in gas-free disks with a wide
variety of surface density distribution of the planetesimal swarm.
Their results indicate that the process of planetary accretion is
fully consistent with the oligarchic growth scenario, {\it i.e.} the
cores initially emerge with a characteristic orbital separation of
$\Delta a_{\rm c} \sim 10 r_{\rm H}$.  In later stages, however,
$\Delta a_{\rm c}$ increases to $\sim 15 r_{\rm H}$, which may be due
to the radial diffusion of the cores (they neglected radial migration
due to disk-planet interaction and did not include the effect of gas
drag on a population of small planetesimals).  Here we adopt $\Delta
a_{\rm c} = 10 r_{\rm H}$.  From eq.~(\ref{eq:sigma_dust}), we
estimate the core masses in protoplanetary systems with
eq.~(\ref{eq:m_iso}) such that
\begin{equation}
M_{\rm c,iso} \simeq
0.16 \eta_{\rm ice}^{3/2} f_{\rm d}^{3/2}
\left(\frac{a}{1\mbox{AU}}\right)^{3/4} M_{\oplus}.
\label{eq:m_iso0}
\end{equation}

If the effect of radial diffusion/migration of the cores is included,
the final core mass could become larger than $M_{\rm c,iso}$.  
In the limiting case that a
migrating core can acquire all the planetesimals interior to its
orbital semi major axis, it would attain a maximum asymptotic mass
\begin{equation}
M_{\rm c,noiso} \sim \pi a^2 \Sigma_{\rm d} 
\simeq
1.2 \left(\frac{\Sigma_{\rm d}}{10 {\rm gcm}^{-2}}\right) 
\left(\frac{a}{1\mbox{AU}}\right)^{2} M_{\oplus}.
\label{eq:m_non_iso}
\end{equation}
Based on the distributions of $\Sigma_{\rm d}$ given by
eq.~(\ref{eq:sigma_dust}), we find
\begin{equation}
M_{\rm c,noiso} \simeq
1.2 \eta_{\rm ice} f_{\rm d} 
\left(\frac{a}{1{\rm AU}}\right)^{1/2} M_{\oplus}.
\label{eq:no_m_iso0}
\end{equation}

Growth beyond isolation is also possible after the gas depletion and
the elimination of its damping effects \citep{Iwasaki02, Kominami02}.
After the gas depletion, the velocity dispersion of the residual cores
grows until they cross each other's orbit.  The final stages of cores'
growth may proceed through giant impacts \citep{Lissauer93}.  
Eventually a few surviving embryos acquire most of the
residual planetesimals and less massive cores during the late
oligarchic-growth stage.  Since their collisional cross section is
reduced to their geometrical surface, the mass of the embryos evolves 
with the growth rate given by eq.~(\ref{eq:mdotembr}), 
\begin{equation}
\begin{array}{ll}
{\displaystyle {d M_{\rm e} \over dt} = {M_{\rm e} \over \tau_{\rm e,acc}} } \; ;
 \\
{\displaystyle \tau_{\rm e,acc} \simeq 4 \times 10^7 
      \eta_{\rm ice}^{-1} f_{\rm d}^{-1} 
      \left( \frac{a}{1{\rm AU}} \right)^{3}
      \left(\frac{M_{\rm e}}{M_{\oplus}} \right)^{1/3} 
      \left(\frac{\rho_{\rm e}}{1{\rm gcm}^{-3}} \right)^{2/3} 
      \left(\frac{M_*}{M_{\odot}} \right)^{1/2} \; {\rm years}}.
\end{array}
\label{eq:emb_acc_rate}
\end{equation}
The asymptotic embryos' masses are given by eq.~(\ref{eq:m_iso}) with
$\Delta a_{\rm c} \sim V_{\rm sruf}/\Omega_{\rm K}$, 
\begin{equation}
M_{\rm e,iso} \simeq
0.52 \eta_{\rm ice}^{3/2} f_{\rm d}^{3/2} 
\left(\frac{a}{1{\rm AU}}\right)^{3/2}
\left(\frac{\rho_{\rm d}}{1{\rm gcm}^{-3}}\right)^{2} M_{\oplus}.
\label{eq:m_e_iso}
\end{equation}

Such a scenario
is consistent with the current model for the origin of the Moon
\citep{Hartmann75, Cameron76}, and the apparently young isotropic age of
the Earth, in comparison with asteroid parent bodies (Halliday et
al.~2000).  Note that even if large-mass cores can be assembled
through giant impacts, they can no longer initiate gas accretion
because the disk gas needs to be severely depleted prior to their
emergence.  The limited gas supply during the late phase of core
growth through giant impacts is also consistent with the intermediate
mass and mostly-ice composition of Uranus and Neptune.

In outer regions, 
the growth of the cores may be terminated before their mass
reaches $M_{\rm e,iso}$ or $M_{\rm c,noiso}$. When their characteristic
surface speed $V_{\rm surf} = \sqrt{G M_{\rm e}/R_{\rm e}}$ exceeds their 
local escape speed from the host star, $V_{\rm escp} = \sqrt {2 G M_*/a}$, the embryos' $90^o$ grazing
scatterings of nearby bodies can lead to the bodies' ejection \citep{Thommes03}.  
In this high $\sigma$ limit, the ratio of collision to ejection
probabilities is $f_{\rm cap} \sim (V_{\rm escp}/V_{\rm surf})^4$.
When $f_{\rm cap}$ become significantly less than unity, the growth of
the embryos is terminated.  Based on previous numerical simulations
\citep{LI97}, we set $f_{\rm cap} \sim 0.1$ such that most
scatterings lead to ejection and the embryos' asymptotic mass becomes
\begin{equation}
{M_{\rm e, sca} \over R_{\rm e}} \sim {2 f_{\rm cap}^{-1/2} M_* \over a}
\end{equation}
where $R_{\rm e}$ is the embryos' physical radius. For an average ice embryo,
with an internal density $\rho_{\rm e} \sim 1$g cm$^{-3}$, the
asymptotic mass of the embryos is
\begin{equation}
M_{\rm e, sca} \sim 2.5 \times 10^2 \left({\rho_{\rm e} \over 1 {\rm g
cm}^{-3}} \right)^{-1/2} f_{\rm cap}^{-3/4} \left( {a \over 1 {\rm AU}}
\right)^{-3/2} M_\oplus \sim 1.4 \times 10^3 \left( {a \over 1 {\rm AU}}
\right)^{-3/2} M_\oplus.
\label{eq:msca}
\end{equation}
This limiting mass at $a \sim 20-30$AU is comparable to those of the cores of Uranus and
Neptune.  Note that this limit only applies to the 
scattering of planetesimals and cores and 
it is not a limiting mass for gas accretion (see \S3).
For notations of masses, time scales, etc., see Table 1. 

In principle, the largest cores may emerge with sufficient asymptotic
masses to initiate gas accretion (see below).  But in many regions,
this large mass can only be attained after the gas is depleted.  Thus,
they are prohibited from evolving into gaseous giant planets.  In this
paper, we consider the formation of the gas giants only if some cores
are able to acquire the critical mass for the onset of rapid gas
accretion before the disk gas is depleted.  We assume that the
coalescence between the cores, through giant impacts after the gas
depletion, do not lead to substantial further growth in the mass of
their gaseous envelopes despite the possibility of substantial 
increases in their core mass.

\subsection{Core growth at different orbital radii}

With the above prescriptions, we estimate, with eq.~(\ref{eq:m_grow}),
the cores' growth rate and asymptotic mass limit.  The initial mass of
the cores is arbitrarily set to be a small value, $M_{\rm c} = 10^{20}$g.  
The choice of an initial value of $M_{\rm c}$ does not affect the results,
since $\tau_{\rm c,acc}$ increases with $M_{\rm c}$.
We plot, in Figure~\ref{fig:core_a_sig}a, the core masses,
after $10^6$ years and $10^7$ years, as a function of $\Sigma_{\rm
d}$ at various locations $a$. For this determination, we assume
that the growth of the cores is truncated with a mass $M_{\rm c,iso}$
(eq.~[\ref{eq:m_iso0}]). We also assume $\Sigma_{\rm g}/\Sigma_{\rm
d}=240$.  A similar plot with the cores' truncation mass set by
$M_{\rm c,noiso}$ (eq.~[\ref{eq:m_non_iso}]) instead of $M_{\rm c,iso}$
is shown in Fig.~\ref{fig:core_a_sig}b.

As shown in eq.~(\ref{eq:m_grow}), the cores' growth is relatively
fast in regions where $\Sigma_{\rm d}$ is high and $a$ is small.
The results in Figures~\ref{fig:core_a_sig} show that for this region
of the parameter space, the cores' growth is terminated with their
asymptotic masses within $10^6$ years.  In contrast, the growth is
relatively slow for regions of parameter space with relatively small
$\Sigma_{\rm d}$ and large $a$ so that $M_{\rm c}(t)$ remains to be
very low in the right lower regions in Figures~\ref{fig:core_a_sig}.
An ``accretion wave'' propagates from the top-left to the lower-right
regions as time passes.  In the very large $\Sigma_{\rm d}$ and
large $a$ regions, $\Delta a_{\rm c} \sim 10 r_{\rm H}$ for $M_{\rm
c,iso}$ exceeds the magnitude of $a$.  Thus, the appropriate
expression for the truncation mass is $M_{\rm c,noiso}$ rather than
$M_{\rm c,iso}$ in these regions.
 
Next, we consider planetary systems with a $\Sigma_{\rm d}$
distribution similar to that derived from a minimum-mass solar nebula
model (see eq.~[\ref{eq:sigma_dust}]).  Figures~\ref{fig:core_evolve}
show some examples of the time evolution of the cores' masses for (a)
$f_{\rm d}=1$ and (b) $f_{\rm d}=10$.  The jumps in the cores'
masses at 2.7AU in these figures reflect a transition in the values of
$\Sigma_{\rm d}$.  These results indicate that in the inner regions
of the disk, the gas depletion time scale is longer than the build-up
time scale for the cores and all the field planetesimals are accreted
onto the isolated cores.  But, for the minimum-mass solar nebula
model (with $f_{\rm d} \simeq 1$), $M_{\rm c,iso}$ at 0.7 and 1 AU
is considerably smaller than the present masses of Venus and Earth
(see eq.~[\ref{eq:m_iso0}]).  These planets could have acquired their
present mass through either giant impacts after depletion of the gas
or the accretion of planetesimals prior to the gas depletion 
if they have diffused or migrated over large
radial distances.  In this case, the cores' asymptotic mass would be
$M_{\rm e,iso}$ (eq.~[\ref{eq:m_e_iso}]) or 
$M_{\rm c,noiso}$ (eq.~[\ref{eq:no_m_iso0}])
which, at 0.7 and 1AU, is comparable to the mass of
the Earth and Venus.

For a more comprehensive parameter study, we show in
Figures~\ref{fig:core_a_fdust}, the dependence of the cores' mass on
$a$ and $f_{\rm d}$ after $10^6$ years and $10^7$ years.
In these calculations, 
$f_{\rm g}$ does not decline and is set to be always equal to
$f_{\rm d}$ since the stages prior to
depletion of the gas are considered. The
magnitude of $\Sigma_{\rm d}$ for the minimum-mass solar nebula
model corresponds to $f_{\rm d} \simeq 1$.  On the gas depletion
time scale $\sim 10^6-10^7$ years, the most massive cores attain a
critical mass ($\sim$ several $M_\oplus$) just outside the ice boundary
(2.7AU).  Although the magnitude of both $M_{\rm c,noiso}$ and 
$M_{\rm c,iso}$ increases with $a$, the expression in eq.~(\ref{eq:m_grow0})
indicates that much beyond $a_{\rm ice}$, the cores do not have sufficient 
time to attain
their asymptotic values prior to the depletion of the disk gas.  In
this case, the formation of gas giants within a minimum-mass solar
nebula is most likely to be initiated just exterior to the ice
boundary.  But around many classical T Tauri stars, the disk mass
inferred from the mm-wave length continuum observation corresponds
$f_{\rm d} \ga 5$ (see \S 1).  The results in
Figs.~\ref{fig:core_a_fdust}a and b suggest that in very massive
protoplanetary disks with $f_{\rm d} \ga 5$, cores can rapidly
attain masses greater than the critical mass
for gas accretion (see next section) even inside the ice boundary.

After the gas depletion, 
$M_{\rm c,iso}$ expands to $M_{\rm c,noiso}$ or $M_{\rm e,iso}$.
$M_{\rm e,sca}$ can also limit embryo growth (see, \S 4.2.3 and
Fig.~\ref{fig:ice_giants}). 
In the calculations in the following sections,
we set the asymptotic masses of cores and embryos as follows:
\begin{itemize}
\item the isolated core model
   \begin{itemize}
   \item min$(M_{\rm c,iso},M_{\rm c,noiso})$, 
         when $f_{\rm g} > 10^{-3}$ (before the gas depletion).
   \item min$(M_{\rm e,iso},M_{\rm c,noiso},M_{\rm e,sca})$, 
         when $f_{\rm g} < 10^{-3}$ (after the gas depletion).  
   \end{itemize}
\item the non isolated core model
   \begin{itemize}
   \item $M_{\rm c,noiso}$,
         when $f_{\rm g} > 10^{-3}$.
   \item min$(M_{\rm c,noiso},M_{\rm e,sca})$ 
         when $f_{\rm g} < 10^{-3}$.
   \end{itemize}
\end{itemize}

\section{Gas accretion}
\label{sec:gas}

Up to now, we only consider the drag effect induced by the gas on the
damping of the planetesimals' and cores' velocity dispersion.  In this
section, we consider the accretion of gas onto sufficiently massive
cores.  The necessary condition for any gas accumulation near the
cores is that their surface escape velocity must exceed the sound
speed of the disk gas.  In regions of the disk where dust can condense
to form planets, the sound speed is less than a few km s$^{-1}$ such
that even lunar-mass cores can attract gas to their proximity.
However, the disk gas first accumulate into a static envelope around
these cores.  In this envelope, the gravity of the cores is balanced
by a pressure gradient which is maintained by the energy release
during the planetesimals bombardment onto the cores.

\subsection{Gas accretion onto cores}
But, when the cores' mass ($M_{\rm c}$) becomes larger than a critical
value ($M_{\rm c,crit}$), the state of quasi hydrostatic equilibrium can
no longer be maintained and a phase of gas accretion onto the
core is initiated \citep{Mizuno80, BP86}.  Thereafter, the
planetesimals and gas accretion proceed concurrently, so that the
planetary growth can no longer be expressed in an analytical form as
that in eq.~(\ref{eq:m_grow0}).  Based on the existing numerical
results we adopt a simple prescription to evaluate $M_{\rm c,crit}$ and
the gas accretion rate, $d M_{\rm p,g} / {dt}$.

The critical core mass depends on the cores' rate of planetesimal
accretion ($\dot{M}_{\rm c}$) and the opacity ($\kappa$) associated
with the disk gas.  Faster accretion and higher opacity (relatively
large $\dot{M}_{\rm c}$ and $\kappa$) result in a warmer planetary
atmosphere and the enhanced pressure gradient stalls the gas accretion
\citep{Stevenson82, Ikoma00}.  Based on a series of numerical models,
\citet{Ikoma00} find that the critical core mass has the following
dependence,
\begin{equation}
M_{\rm c,crit} \simeq 
10 \left( \frac{\dot{M}_{\rm c}}{10^{-6}M_{\oplus} {\rm year}^{-1}}\right)^{0.2-0.3}
\left( \frac{\kappa}{1{\rm cm^2 g^{-1}}}\right)^{0.2-0.3}M_{\oplus}.
\label{eq:crit_core_mass}
\end{equation}
We adopt here a power index $1/4$ for the dependence on $\dot{M}_{\rm c}$.
If the dust particles have a similar distribution and abundance as the
interstellar medium, $\kappa \sim 1 {\rm cm}^2 {\rm g}^{-1}$.  The actual
magnitude of $\kappa$ is uncertain since the dust particles' size
distribution and the fraction of heavy element in the gas and solid
phases are poorly known \citep{Podolak03}.  For computational simplicity, we neglect,
the dependence of $M_{\rm c,crit}$ on $\kappa$ in our prescription below.

The gaseous envelope of proto giant planets contracts on a Kelvin
Helmholtz time scale when their masses $M_{\rm p}$ (including both their
solid cores and gaseous envelope) become larger than $M_{\rm c,crit}$.  
\citet{Ikoma00} derived through numerical calculation that
\begin{equation}
\tau_{\rm KH} \simeq 10^b \left( \frac{M_{\rm p}}{M_{\oplus}}
\right)^{-c}
\left( \frac{\kappa}{1{\rm gcm}^{-2}}\right)
\; {\rm years}
\label{eq:gas_acc_rate1}
\end{equation}
with power index $b \simeq 8$ and $c \simeq 2.5$.  However, with
different opacity table, the values of $b$ and $c$ can be changed to be $\simeq
10$ and $\simeq 3.5$ respectively (Ikoma, private communication).
\citet{Bryden00b} obtained $b \simeq 10, c \simeq 3.0$ by fitting the
result of \citet{P96}.  We adopt here $b = 9$ and $c = 3.0$ and also
neglect the dependence on $\kappa$ in our prescription below.

The contraction of the gas envelope reduces the local pressure
gradient.  Consequently, the gas from the surrounding disk flows toward
the cores to replenish the envelope at a rate
\begin{equation}
\frac{dM_{\rm p,g}}{dt} \simeq \frac{M_{\rm p}}{\tau_{\rm KH} } 
\label{eq:mgsdot}
\end{equation}
This rate is well below the Bondi accretion rate \citep{Bondi52} when
$M_{\rm p} \sim M_{\rm c,crit}$ so the gas accretion rate is insensitive
to the boundary condition in the disk.  The magnitude of $dM_{\rm
p,g}/{dt}$ due to the Kelvin Helmholtz contraction increases rapidly
with $M_{\rm p}$.  In principle, $dM_{\rm p,g}/{dt}$ is limited by
the Bondi accretion rate \citep{Bondi52}, the Keplerian shear in the
gas inflow \citep{Tanigawa00}, mass diffusion rate through the disk
\citet{Bryden99}, and the availability of gas near the orbit of the
accreting planet.  But, these effects become important only when
$M_{\rm p}$ has already become a significant fraction of Jupiter's
mass $M_{\rm J}$.  For $M_{\rm p} \sim M_{\rm J}$, 
$\tau_{\rm KH}$ is reduced to the
dynamical free-fall time scale and the growth time scale $M_{\rm p}/
(dM_{\rm p,g}/{dt})$ for Bondi accretion is reduced to $10^3$ years.
Thus, the total gas accretion time for the protoplanets to attain
their asymptotic masses is essentially determined by $\tau_{\rm KH}$
at $M_{\rm p} \sim M_{\rm c,crit}$.

Note that when $M_{\rm c}$ reaches $M_{\rm c,iso}$, $\dot{M}_{\rm c}$
is reduced to zero.  Consequently, $M_{\rm c,crit}$ also vanishes and
gas accretion can start for any value of $M_{\rm c}$.  However, if
\begin{equation}
M_{\rm c} \la M_{\rm c,KH} \equiv M_{\rm p} (\tau_{\rm KH} = \tau_{\rm disk})
\simeq 4.6 \left( {\tau_{\rm disk} \over 10^7 {\rm year}} \right)^{-1/3}
M_{\oplus}, 
\label{eq:mkh}
\end{equation}
the gas accretion rate is so small (eq.~[\ref{eq:mgsdot}]) that $M_{\rm
p}$ is not significantly increased by the gas accretion during the
first $10^7$ years after they were formed.  In order for the gas
giants to actually form and acquire $M_{\rm p} \gg M_{\rm c,crit}$,
$M_{\rm c}$ must be larger than $M_{\rm c,KH}$ in addition to
the requirement of $M_{\rm c} > M_{\rm c,crit}$.  Both of conditions can
be satisfied for $M_{\rm c} \ga 4 M_{\oplus}$, if the cores' rate of
planetesimal accretion is significantly reduced from that in
eq.~(\ref{eq:mdot}).
 
\subsection{A prescription for the growth of gas giant planets}
Based on the above considerations, we introduce the following simple
prescription for the purpose of numerically integrating the formation
and early evolution of the gaseous giant planets.  These approximation
captures the essence of the planet formation process.

\begin{itemize}
\item Proto planetary growth due to planetesimal accretion is computed 
      with the following rate equation,
      \begin{equation}
      \frac{dM_{\rm p,pl}}{dt} = \frac{M_{\rm p}}{\tau_{\rm c,acc}}
      = \dot M_{\rm c} (M_{\rm c} = M_{\rm p}), 
      \label{eq:mpdot_cal}
      \end{equation}
      where $\tau_{\rm c,acc}$ is given by eq.~(\ref{eq:pl_acc_rate}).
      The value of $\dot{M}_{\rm c}$ is obtained from eq.~(\ref{eq:mdot}) with
      the core mass $M_{\rm c}$ being replaced by the protoplanet's total
      mass $M_{\rm p}$ which also includes the mass of the gaseous envelope. 
\item During the growth of the cores, we check whether $M_{\rm p}$ is
      less than the critical core mass for the onset of gas accretion.
      Based on the discussions in the previous subsection, the magnitude of
      $M_{\rm c,crit}$ is approximated with a simplified version of
      eq.~(\ref{eq:crit_core_mass}) such that 
      \begin{equation} 
      M_{\rm c,crit} \simeq 10
      \left( \frac{\dot{M}_{\rm c} (M_{\rm c} = M_{\rm p})}
      {10^{-6} M_{\oplus} {\rm year}^{-1}}\right)^{1/4} M_{\oplus}.  
      \label{eq:crit_core_mass_we_use}
      \end{equation} 
      In principle, as the field planetesimals are being accreted onto 
      the cores, their surface density in the feeding zone declines, 
      so that both $\dot{M}_{\rm c}$ and $M_{\rm c,crit}$ decrease.
      To take this effect into account, we use time-dependent values
      of $\dot{M}_{\rm c}$ calculated in (\ref{eq:mpdot_cal}) and set 
      $M_{\rm c,crit}=0$ for cores with 
      $M_{\rm c}>M_{\rm c,iso}$ or $M_{\rm c,noiso}$.
\item When $M_{\rm c}$ exceeds $M_{\rm c,crit}$, gas
      accretion onto a core is set to start. The gas accretion rate is
      regulated by the efficiency of heat transfer and it is 
      approximated by a simplified version of eqs.~(\ref{eq:gas_acc_rate1}) 
      and (\ref{eq:mgsdot}) such that
      \begin{equation}
      \begin{array}{l}
     {\displaystyle       
      \frac{dM_{\rm p,g}}{dt} \simeq \frac{M_{\rm p}}{\tau_{\rm KH}} \, ;} \\
     {\displaystyle       
      \tau_{\rm KH} \simeq 10^9 \left( \frac{M_{\rm p}}{M_{\oplus}}
      \right)^{-3} \mbox{ years}}.
      \label{eq:gas_acc_rate_we_use}
      \end{array}
      \end{equation}
      The planets' mass $M_{\rm p}$ includes both their gaseous envelope
      and cores.  Due to the expansion of the growing planets' feed zones, 
      planetesimal accretion onto the cores continues after the gas accretion 
      has started.  But, we neglect the collisions between cores until after 
      the disk gas is depleted as we have discussed in the last section.
\item The above gas accretion rate does not explicitly depend on the ambient 
      conditions.  The approximation is appropriate provided there is an 
      adequate supply of the disk gas. However, the disk gas may be depleted
      either globally through processes such as the viscous evolution, outflow,
      and photo evaporation or locally through gap formation.  These effects
      are taken into account in the determination of the asymptotic mass of the
      giant planets (see the next subsection) but not in their accretion rates.
\item In the limit that the growth of the gas giant planets is terminated 
      by gap formation, we also consider their dynamical evolution with or 
      without the type-II orbital migration (see the next section).  
\end{itemize}

\subsection{Termination of gas accretion}

Equation (\ref{eq:gas_acc_rate_we_use}) shows that the gas accretion
rate rapidly increases with $M_{\rm p}$.  However, there are several
asymptotic limits to the unimpeded runaway gas accretion process.  These
limits arise due to the local and global depletion of disk gas.

\subsubsection{Local gas depletion}
In an inviscid disk, gas accretion onto the cores is terminated when
they consume all the gas in their feeding zone and attain an isolation
mass.  The isolation mass for the gas giants can be obtained by
replacing $\Sigma_{\rm d}$ in eq.~(\ref{eq:m_iso}) with
$\Sigma_{\rm g}$ such as
\begin{equation}
M_{\rm g,iso} \simeq
50 \left(\frac{\Sigma_{\rm g}}{2.4 \times 10^3 \mbox{gcm}^{-2}}\right)^{3/2}
\left(\frac{a}{1\mbox{AU}}\right)^{3}
\left(\frac{\Delta a_{\rm g}}{2r_{\rm H}} \right)^{3/2} 
\left(\frac{M_*}{M_{\odot}} \right)^{-1/2} 
M_{\oplus},
\label{eq:m_gas_iso}
\end{equation}
where the Hill (Roche lobe) radius $r_{\rm H}$ is evaluated with
$M_{\rm p}$.  For a fiducial magnitude of $M_{\rm g,iso}$, we used the
width of gas feeding zone $\Delta a_{\rm g} = 2r_{\rm H}$ to be a
nominal value.  Since it flows normally on circular orbits, gas which
is much beyond $\Delta a_{\rm g}$ cannot reach the accreting cores
without modifying their specific angular momentum.  Using the
$\Sigma_{\rm g}$ distribution in eq.~({\ref{eq:sigma_g}), we find
\begin{equation}
M_{\rm g,iso} \simeq 50 f_{\rm g}^{3/2}
\left(\frac{a}{1{\rm AU}} \right)^{3/4} 
\left(\frac{\Delta a_{\rm g}}{2r_{\rm H}} \right)^{3/2} 
\left(\frac{M_*}{M_{\odot}} \right)^{-1/2} M_{\oplus},
\label{eq:m_gas_iso0}
\end{equation}
which is comparable to $M_{\rm J}$ at the present location of Jupiter.
Since $\Sigma_{\rm g}$ and equivalently $f_{\rm g}$ decrease with time,
$M_{\rm g, iso}$ also decreases with time.
$dM_{\rm p,g}/dt$ is set to be zero when $M_{\rm p}$ becomes
larger than $M_{\rm g, iso}$.

However, the observed accretion flow from protostellar disks to their
host stars \citep{Hartmann98} requires mass and angular momentum
transfer throughout the disks.  In a viscously evolving disk, gas can
continuously diffuse into the feeding zone until the growing planet
attains a mass \citep{LP85},
\begin{equation}
M_{\rm g,vis} \simeq \frac{40 \nu}{a \Omega_{\rm K}} 
M_* \simeq 40 \alpha \left(\frac{h}{a}\right)^2 M_*
\simeq 30 \left(\frac{\alpha}{10^{-3}}\right)
\left(\frac{a}{1{\rm AU}}\right)^{1/2}
\left(\frac{M_*}{M_{\odot}}\right)  M_{\oplus},
\label{eq:m_gas_vis}
\end{equation}
where we used eq.~(\ref{eq:h_a}) and $\alpha$-prescription for the
effective viscosity $\nu$ (Shakura \& Sunyaev 1973), in which
$\nu = \alpha h^2 \Omega_{\rm K}$ where $\alpha$ is a dimensionless
parameter.  For $M_{\rm p} >
M_{\rm g,vis}$, an embedded planet induces a strong tidal torque to
open a gap near its orbit, provided its Hill (Roche lobe) radius is
larger than a critical value $r_{\rm H,c} = h$ \citep{LP93, Bryden00a}.  
We find that $r_{\rm H} = r_{\rm H,c}$ when
\begin{equation}
M_{\rm g,th} 
\simeq 3.8 \times 10^{-4} \left(\frac{a}{1{\rm AU}}\right)^{3/4}
M_*
\simeq 1.2 \times 10^{2} \left(\frac{a}{1{\rm AU}}\right)^{3/4}
\left(\frac{M_*}{M_{\odot}}\right) M_{\oplus}.
\label{eq:m_gas_therm}
\end{equation}
Because $M_{\rm g,th}$ is generally larger than $M_{\rm g,vis}$
except for some very large $\alpha$ cases, gas inflow onto the planet
would actually be halted, when $M_{\rm p}$ reaches $M_{\rm g,th}$.
In some previous hydrodynamic simulations, embedded protoplanets
continue to accrete after the above condition is satisfied albeit at a
reduced rate \citep{AL96}.  In oder to take this possibility into account,
we also modify, in one set of calculations, the truncation condition
to $r_{\rm H,c} = 1.5 h$.

\subsubsection{Global gas depletion}

Before their $M_{\rm p}$ increases to the isolation mass, 
$M_{\rm gas,iso}$, protoplanets can accrete at the rate given in
eq.~(\ref{eq:gas_acc_rate_we_use}).  But, for $M_{\rm p} > M_{\rm
gas,iso}$, additional disk material can no longer be brought into the
feeding zone by the expansion of its boundary.  Nevertheless, in a
viscously evolving disk, the feeding zone may be replenished.  
We also consider the extreme case that the
protoplanets' tidal interaction does not impinge the flow of the gas
from the disk to their feeding zone.

Although gas diffusion can replenish the feeding zone, the rate of gas
accretion onto the protoplanet is limited to
\begin{equation}
{d M_{\rm p, g} \over dt} = {\rm min} \left({M_{\rm p} \over \tau_{\rm KH}},
\dot M_{\rm disk} \right)
\end{equation}
where the mass transfer rate in the disk,
\begin{equation}
\dot M_{\rm disk} = 2 \pi \Sigma_{\rm g} a V_{\rm disk},
\end{equation}
the gas diffusion speed is determined by the efficiency 
of viscous transport 
\begin{equation}
V_{\rm disk} \simeq - {3 \nu \over a} \left( {\partial {\rm ln}\, 
(\Sigma_{\rm g} \nu) \over \partial {\rm ln}\, a} + {1 \over 2} \right).
\end{equation}
With the $\alpha$ prescription for a steady disk, $V_{\rm disk} 
\sim - \alpha h^2 \Omega_{\rm K}/a $.

The magnitude of $\alpha$ in the planet formation region is uncertain.  
We assume that the mass transfer rate throughout the disk
is comparable to an observational inferred accretion rate from
protostellar disks onto classical T Tauri stars \citep{Hartmann98} 
which can be fitted by 
\begin{equation}
\dot M_{\rm disk} \sim \dot M_0 (t / \tau_{\rm disk})^{-1.5} 
\label{eq:leemdot}
\end{equation}
where $\dot M_0 = 10^{-4} M_\oplus$ year$^{-1}$ for $\tau_{\rm disk} = 10^7$ yr
\citep{Calvet00}.  The actual data have an order of magnitude spread over
this fit.  The accretion rate onto the planets is limited by the gas 
replenishment rate of their feeding zone, $\dot M_{\rm disk}$, when their 
\begin{equation}
\frac{M_{\rm p}}{\tau_{\rm KH}} > \dot M_{\rm disk}
\label{eq:mdotlim}
\end{equation}
or equivalently when their $M_{\rm p}$ becomes larger than
\begin{equation}
M_{\rm g,sup} (t) \simeq 20 \left({t \over 10^7 {\rm years}} \right)^{-3/8} 
M_\oplus
\end{equation}
(see eqs.~[\ref{eq:gas_acc_rate1}] and [\ref{eq:mdotlim}]).  The limiting
replenishment rate only modifies $d M_{\rm p,gas}/dt$ for protoplanets
more massive than both $M_{\rm g,sup}$ and $M_{\rm g,iso}$.  From
eqs.~(\ref{eq:m_gas_iso0}) and (\ref{eq:fgas_decay}), 
we find that these conditions are satisfied
at the advanced stages of disk evolution when the surface density of
the disk is significantly depleted.

At any given location, the emergence of cores which can dynamically
accrete gas occurs at an epoch
\begin{equation}
\tau_{\rm onset} = 3 \tau_{\rm c, acc} (M_{\rm c} = M_{\rm c,crit})
\sim 8 \times 10^5 \eta_{\rm ice}^{-1} f_{\rm d}^{-7/5}
\left({a \over 1 {\rm AU}} \right)^{27/10} 
\left({M_{\rm c,crit} \over 10 M_\oplus} \right) ^{1/3} {\rm years}
\end{equation} 
for $M_\ast = 1 M_\odot$.
According to eq.~(\ref{eq:leemdot}), the amount of mass can be supplied
into the feeding zone for $t > \tau_{\rm onset}$ is
\begin{equation}
M_{\rm g, feed} = \int_{\tau_{\rm onset}}^\infty \dot M_{\rm disk} (t)
dt \simeq 1.5 \dot M_{\rm disk} (t=\tau_{\rm onset})
\tau_{\rm disk} \simeq 60 \left( {\eta_{\rm ice} \over 4} 
\right)^{3/2} f_{\rm d}^{21/10}
\left({a \over 10 {\rm AU}} \right)^{-81/20} \left({
\tau_{\rm disk} \over 10^7 {\rm years} }\right)^{5/2} M_\oplus.
\label{eq:m_gas_feed}
\end{equation}
If such supply is considered, the isolation mass 
can increase from $M_{\rm g, iso}$ to $M_{\rm g, iso} + M_{\rm g, feed}$. 
However, at large $a$, $M_{\rm g, feed}$ is not a large enough.
Since sufficiently massive cores (with $M_{\rm c} > M_{\rm c,crit}$)
take longer to form in the outer regions of the disk, the mass
transfer rate can limit the amount of material supplied to the feeding
zone, $M_{\rm g, feed}$, to less than the mass of Jupiter.

The gas accretion is ultimately limited by the amount of residual gas
in the entire disk.  For our disk model, the maximum available mass is
\begin{equation}
M_{\rm g,noiso} \sim \pi a^2 \Sigma_{\rm g} \simeq
290 \left(\frac{\Sigma_{\rm g}}{2.4 \times 10^3 \mbox{gcm}^{-2}}\right)
\left(\frac{a}{1\mbox{AU}}\right)^{2} M_{\oplus} 
\simeq  290 f_{\rm g} \left(\frac{a}{1\mbox{AU}}\right)^{1/2} M_{\oplus}.
\label{eq:m_gas_non_iso}
\end{equation}
When $M_{\rm g,noiso} < M_{\rm g,th}$, $M_{\rm p}$ is truncated by
$M_{\rm g,noiso}$ rather than $M_{\rm g,th}$.  This condition
occurs when $f_{\rm g} < 3 (a/1{\rm AU})^{1/4}
(M_*/M_{\oplus})^{5/2}$.  
Similar to $M_{\rm g,iso}$, $M_{\rm g,noiso}$
explicitly depends on the magnitude of $f_{\rm g}$, {\it i.e.}, the
amount of residual gas in the disk.

As we have already indicated in \S 2 that the gas content are
observed to decline concurrently with the dust (but not the planetesimal)
content in protostellar disks.  Since the near IR signature for dust
in the inner region declines together with the mm indicators of dust
in the outer regions of protostellar disks \citep{Duvert00}, we assume that
the residual gas in the these disks is depleted globally in a self
similar manner.  We introduce a prescription to approximate the
evolution of $\Sigma_{\rm g}$ during the depletion phase of the disk
such that
\begin{equation}
\Sigma_{\rm g} \simeq \Sigma_{{\rm g},0} \exp\left(
-\frac{t}{\tau_{\rm disk}} \right),
\label{eq:gas_decay}
\end{equation}
where $\Sigma_{\rm g,0}$ is an initial distribution
(eq.~[\ref{eq:sigma_g}]) and $\tau_{\rm disk}$ is gas depletion time
scale.  Equivalently, $f_{\rm g}$ decreases with time from its
initial value $f_{\rm g,0}$ as
\begin{equation}
f_{\rm g} \simeq f_{{\rm g},0} 
\exp\left( -\frac{t}{\tau_{\rm disk}} \right).
\label{eq:fgas_decay}
\end{equation}
Based on the observed properties of protostellar disks 
\citep{BS96,Wyatt03},
we set $\tau_{\rm disk} = 10^{7}$ years.

In order to limit the extent of the parameter studies, we also assume
here that $f_{\rm g,0} = f_{\rm d}$, that is, the gas-dust ratio is
initially equal to the Solar composition.  We denote both $f_{\rm g,0}$ 
and $f_{\rm d}$ to be $f_{\rm disk}$.  We consider the cases
with $f_{\rm disk}=0.1-30$ as mentioned in the last section.  We do
not vary the value of $f_{\rm disk}$ during the cores' growth but it
is locally set to zero when the cores have reach their asymptotic
masses.

Based on these prescriptions, we set an asymptotic mass for the gas
giants.  We assume that the gas accretion onto it is terminated when a
protogiant planets either consumes all the nearby residual gas in the
disk or has acquired a sufficiently large mass to open up a gap near
its orbit.

\section{Nascent Orbital Distribution of Potoplanets}

In this section, we consider dynamical properties such as the
mass-semi major axis distribution which planets are formed with. This
distribution evolves with the orbital migration of the gas giant
planets.  We will discuss effect of orbital migration in the next
section.

\subsection{Conditions for the emergence of gaseous and ice giant planets}  
\label{sec:sec41}
We compute a series of models with $\tau_{\rm disk} = 10^7$ years.  We
determine the asymptotic (at $t=10^9$ years) magnitude of 
$M_{\rm p}$, including that of the gaseous envelope if any, as a function of
$f_{\rm disk}$ ($=f_{\rm d}=f_{\rm g,0}$) and 
$a$ (see Figures~\ref{fig:core_gas_iso}).  For these
models, we adopted the isolated core model (\S 2.5).
We adopt five different prescriptions
for the planets' asymptotic mass after their gas accretion is
truncated: (a) $M_{\rm g,iso}$, (b) $M_{\rm g,noniso}$, (c) $M_{\rm gas,vis}$, 
(d) $M_{\rm g,th}$, and (e) $M_{\rm g,iso} + M_{\rm g,feed}$.
For the evaluation of the asymptotic mass of the gas
giants, we adopt $\Delta a_{\rm g} = 2r_{\rm H}$ in (a) and (e), $M_* = 1
M_{\odot}$ in (c) and (d), and $\alpha = 10^{-3}$ in (c).  
Since the result of (e) is almost the same as that of (d), it is omitted.
There are some uncertainties in the amount of gas diffusion to the planet's
feeding zone.  These five upper mass limits provide a range in the
asymptotic mass of the planets.  In Figs.~7, we show the results
of a second series of models computed with the non isolated core model 
(\S 2.5).  All other
parameters are identical to the first series of models.

For disks with a uniform $f_{\rm disk} \sim 1-3$, there are three
distinct regions in Figures~\ref{fig:core_gas_iso}.  Similar regions
also exist in Figures~\ref{fig:core_gas_noiso} for disk with uniform
$f_{\rm disk} \sim 0.3-1$.  In the inner ($a<a_{\rm ice}$) regions, the
asymptotic mass of the cores $M_{\rm c}$ is smaller than
$4M_{\oplus}$, which is the minimum mass ($M_{\rm c,KH}$) needed to
undergo a significant amount of gas accretion within $10^7$ years.
Since gas is depleted on this time scale, gas giants can not form in
these region.  Although after gas depletion, the cores' mass can
exceed several $M_{\oplus}$ through giant impacts between them, they
can no longer accrete gas.  These massive cores evolve into
terrestrial planets similar to the Earth and the mass distinction
between the terrestrial planets and gas giants is preserved.

In the outer regions of the disk where $a \ga 10-20$AU, the cores'
planetesimal accretion rate is very low.  Although the asymptotic mass
limit for the cores is large compared with several $M_\oplus$, there is
inadequate amount of time for the cores to attain a significant amount
of mass prior to the depletion of the disk gas.  Nevertheless, low
mass cores can slowly accrete gas because, for very low planetesimal
accretion rates $\dot M_{\rm c}$, the values of $M_{\rm c,crit}$ is also
small.  But, in the low-$\dot M_{\rm c}$ limit, the gas disk is
severely depleted by the time $M_{\rm c}$ becomes comparable to
$M_{\rm c,KH}$.  In advanced stages of gas depletion, the magnitude of
both $M_{\rm g,iso}$ and $M_{\rm g,noiso}$ becomes smaller than
$M_{\rm c}$ and the cores cannot acquire a massive gaseous envelope.
This limited supply of residual disk gas may have quenched the ability
of Uranus and Neptune to attain a massive envelope ($\la 2 M_\oplus$)
despite their present ice-core mass being $\ga 15M_{\oplus}$\citep{P96}.
We refer to these heavy element enhanced, late comers with modest gaseous
envelopes as ice giant planets. 

In the intermediate region slightly exterior to the ice boundaries, an
abrupt increase in the cores' asymptotic mass allows the cores to
initiate the onset of rapid gas accretion. Gas giants with $M_{\rm p}
\ga 100M_{\oplus}$ are preferentially formed in this bright region in
Figures~\ref{fig:core_gas_iso} and \ref{fig:core_gas_noiso}.  The
boundaries of the region where gas giants can form are set by the
requirements that: (a) $M_{\rm c} \ga 4M_{\oplus}$ and (b) $3
\tau_{\rm c,acc} \la \tau_{\rm disk}$.  The results in
Figures~\ref{fig:core_gas_iso} and \ref{fig:core_gas_noiso} indicate
that conditions (a) and (b) are satisfied, for $f_{\rm disk} \ga 1$, at
$a_{\rm tg}$ (the "rocky/gas-giant planets' boundary")
$\ga 3$AU and $a_{\rm gi}$ (the "gas/ice
giants' boundary") $\la 10-20$AU.  {\it Terrestrial planets
form interior to $a_{\rm tg}$ and ice giants form exterior to $a_{\rm
gi}$.}  The values of of $a_{\rm tg}$ and $a_{\rm gi}$ are 
expressed in eqs.~(\ref{eq:atg}) and (\ref{eq:agi}) below.

We now consider $a_{\rm tg}$ and $a_{\rm gi}$ 
for more general values of $f_{\rm disk}$.
Transition from cores to gas giants requires gas accretion and can
only occur prior to gas depletion.  But, in the presence of gas, the
core growth is limited by their asymptotic masses.  If the asymptotic
mass is determined by the isolation process, it would become $M_{\rm
c,iso}$.  In this case, condition (a) can only be satisfied if $M_{\rm
c,iso} > 4 M_\oplus$.  From eq.~(\ref{eq:m_iso0}), this necessary
condition corresponds to
\begin{equation}
f_{\rm disk} \ga f_{\rm tg} \simeq 8 \eta_{\rm ice}^{-1} \left( 
\frac{a}{1{\rm AU}} \right)^{-1/2}.
\label{eq:giants_form_cond1}
\end{equation}
Similarly, in the non isolated limit, the condition that $M_{\rm
c,noiso} > 4 M_\oplus$  corresponds to 
\begin{equation}
f_{\rm disk} \ga f_{\rm tg} \simeq 3 \eta_{\rm ice}^{-1} \left( 
\frac{a}{1{\rm AU}} \right)^{-1/2}
\label{eq:giants_form_cond2}
\end{equation}
(see eq.~[\ref{eq:no_m_iso0}]). 
These estimates agree with the terrestrial/gas-giant planets'
boundary in Figures~\ref{fig:core_gas_iso} and
\ref{fig:core_gas_noiso}, respectively.  In comparison with the results
in the first series, the emergence of the more massive cores in the
second series of simulations enhances the probability and speeds up
the onset of gas giant formation for modest values of $a$.

In condition (b) ($3 \tau_{\rm c,acc} \la \tau_{\rm disk}$), the
quantity $3 \tau_{\rm c,acc}$ is the actual growth time scale for the
cores (see eq.~[\ref{eq:pl_acc_rate}]).  From
eq.~(\ref{eq:pl_acc_rate}), the condition for $3\tau_{\rm c,acc}
\simeq \tau_{\rm disk} $ corresponds to
\begin{equation}
f_{\rm disk} \ga f_{\rm gi} \simeq 4 
\left(\frac{\eta_{\rm ice}}{4}\right)^{-5/7} 
\left(\frac{M_{\rm p}}{4 M_{\oplus}}\right)^{5/21} 
\left(\frac{\tau_{\rm dep}}{10^7 {\rm years}}\right)^{5/7} 
\left( \frac{a}{10{\rm AU}} \right)^{27/14}.
\label{eq:giants_form_cond3}
\end{equation}
This condition does not depend on whether the asymptotic mass of the
cores is limited by $M_{\rm c,iso}$ or $M_{\rm c,noiso}$.  For
$\eta_{\rm ice}=4$, $M_{\rm p}=4 M_{\oplus}$, and $\tau_{\rm disk} =
10^7$ years, this equation indicates that $f_{\rm gi} \simeq 4
(a/10{\rm AU})^{1/2}$, which also agrees well with the gas/ice giants'
boundary in Figures~\ref{fig:core_gas_iso} and
\ref{fig:core_gas_noiso}.

As the analytical derivation shows, the two boundaries separating the
three regions nurture the formation of terrestrial planets, gas and ice
giants and they do not depend on conditions of truncation of gas
accretion.  In the outer regions of the disk, the gas/ice giants'
boundary is also insensitive to the asymptotic mass of the cores.  At
large $a$, the barrier for reaching $M_{\rm c,crit}$ is imposed by the
persistence of disk gas on time scales $\tau_{\rm disk}$ rather than
the asymptotic limit of the core's mass.

For disks with arbitrary values of $f_{\rm disk}$, the boundaries 
\begin{equation}
a_{\rm tg}= 
\left\{
\begin{array}{ll}
64 f_{\rm disk}^{-2} \eta_{\rm ice}\; {\rm AU} & (\mbox{the isolated core case)} \\
 9 f_{\rm disk}^{-2} \eta_{\rm ice}\; {\rm AU} & (\mbox{the non isolated core case)} ,
\end{array}
\right.
\label{eq:atg}
\end{equation}
and
\begin{equation}
a_{\rm gi} = 5 f_{\rm disk}^{14/27}
\left( {\eta_{\rm ice} \over 4} \right)^{10/27} 
\left( {M_{\rm p} \over 4 M_{\oplus} } \right)^{-10/81} 
\left( {\tau_{\rm dep} \over 10^7 {\rm years} } \right)^{-10/27} \; {\rm AU}
\label{eq:agi}
\end{equation}
are set by the conditions $f_{\rm disk} = f_{\rm tg}$ and $f_{\rm gi}$
respectively.  In massive disks with considerable heavy elemental
contents, $f_{\rm disk}$ may be considerably larger than unity.  In these
massive disks, $a_{\rm tg}$ is less than the ice boundary.  In that
case, it is possible to form gas giants interior to the ice boundary.
In disks with $f_{\rm disk}$ smaller than a critical value
\begin{equation}
f_{\rm disk, min} = 0.9 \; {\rm or} \; 0.4 
\left( {\eta_{\rm ice} \over 4} \right)^{-64/68}
\left( {M_{\rm p} \over 4 M_\oplus} \right)^{5/102} \left( {\tau_{\rm dep}
\over 10^7 {\rm years}} \right)^{5/34},
\label{eq:fcrit}
\end{equation}
$a_{\rm gi} < a_{\rm tg}$ (the factors 0.9 and 0.4 correspond to
the isolated and non isolated cases, respectively).
These results
are in agreement with the numerical results presented in
Figures~\ref{fig:core_gas_iso} and ~\ref{fig:core_gas_noiso}.  

Since cores must attain at least $4M_\oplus$ before they can rapidly accrete 
gas, {\it in metal deficit
disks with $f_{\rm disk} < f_{\rm disk, min}$, the gas giants cannot form}.
For a minimum mass solar nebula model, $f_{\rm disk} =1$ and is 
$ < f_{\rm tg}$ inside the ice line and $> f_{\rm gi}$ outside the
ice line such that gas giants can only form at large radii.  Since
$f_{\rm disk, min}$ is smaller beyond the ice line, it sets a more
stringent condition with $\eta_{\rm ice} =4$.  
They are also consistent with the apparent paucity of metal deficit
stars with planets, e.g., \citep{Gonzalez97,Santos01,Murray02} 
and (D. Fischer, private communication).

In the gas giant formation domain, the total mass of the planets
increases rapidly through runaway gas accretion after it exceeds
$M_{\rm c,KH} \sim 4 M_\oplus$.  Gas accretion continues until the gas
is depleted either globally or locally near the planets' orbits.
Since the planets' growth time scale from the onset of rapid gas
accretion to their asymptotic mass is relatively short compared with
the disk depletion time scale, planets with $M_{\rm p} \sim 10-100
M_{\oplus}$ occupy a very limited domain interior to $a \la 3$AU, as
shown in Figures~\ref{fig:core_gas_iso} and \ref{fig:core_gas_noiso}
(with the exception of the case where $M_{\rm p}$ is limited by $M_{\rm g,
vis}$).  In outer regions ($a \ga 3-10$AU), however, ice giants in
such mass ranges can be formed.  Their cores attain sufficient mass to
accrete gas only after the gas has already been depleted severely.

\subsection{Prediction for mass and semi major axis distribution}  

Using the same theoretical model of core growth and gas accretion, we
performed Monte Carlo calculations to produce a theoretical prediction
for the $M_{\rm p}-a$ distribution of extra solar planets to compare
it with observations (Figure~\ref{fig:obs}).  Planets are assumed to
form with equal probability per interval of $\log a$.  
We assume the $f_{\rm disk}$ distribution as in
Fig.~\ref{fig:f_disk_dist} to be consistent with
observational data \citep{BS96, Wyatt03}.  
This is a gaussian distribution in terms of $\log_{10} f_{\rm disk}$
with a center at $\log_{10} f_{\rm disk} = 0.25$ and dispersion of 1. 

We omit the high $f_{\rm disk}$ tail at $f_{\rm disk} > 30$, since
such heavy disks are self gravitationally unstable in outer regions.
The gravitational stability parameter \citep{Toomre64} for our disk models is
\begin{equation}
Q={c_s \Omega_{\rm K} \over \pi G \Sigma_{\rm g} } \simeq
40 f_{\rm disk} ^{-1} \left( {a \over 1 {\rm AU}} \right)^{-1/4} .
\label{eq:toomre_q}
\end{equation}
The mass transfer associated with the gravitational instability 
would rapidly decrease local surface density $\Sigma_{\rm g}$ to 
a value such that $Q > 1$ \citep{LP90, Nakamoto94}. 
Surface density in outer regions ($a \ga 10$AU) in 
massive disks ($f_{\rm disk} \ga 20$) may be adjusted.
In principle, we should use the adjusted surface density
distribution.  However, since most of giant planets found in our model 
are formed at $a \la 10$AU, we adopt a simple cut-off at
$f_{\rm disk} = 30$.
We also examined a more artificial distribution in which
$f_{\rm disk}$ is uniform in log scale from 0.1 to 30.
But, the results are almost the same.

$\tau_{\rm disk}$ is distributed from $10^6$ to $10^7$ years 
uniformly in log scale, which is also consistent with
observational data \citep{BS96, Wyatt03}.  
The range of stellar mass
is also considered, so that $a_{\rm ice}$ is changed as
eq.~(\ref{eq:a_ice}).  The location of $a_{\rm ice}$ is essential for
the formation of gas giants, as shown before.  In order to directly
compare with the observed data, we adopt a similar selection criterion
for $M_*$ in the range of $0.7-1.4M_{\odot}$.  Correspondingly,
$a_{\rm ice} = 1.3-5.3$AU.  We also assume a uniform distribution in
log scale for $M_*$.  In each run, 10,000 initial conditions are taken.

We carry out calculations for planetary formation in disks with an
evolving gas content.  In these models, the cores' accretion is
truncated when their masses have attained asymptotic values of $M_{\rm
c,iso}$ (the results with $M_{\rm c,noiso}$ are similar).  These
results are shown in Figure~\ref{fig:ma}.  Three truncation
asymptotic masses resulting from the truncation of gas accretion are
considered: (a) $M_{\rm g,iso}$, (b) and (c) $M_{\rm g,th}$, and (d)
$M_{\rm g,vis}$.  We adopt $\Delta a_{\rm g} = 2r_{\rm H}$ in (a), the
critical Hill's radius $r_{\rm H,c}$ being $h$ and $1.5h$ in (b) and
(c), respectively, and $\alpha = 10^{-3}$ in (d).

In Figure~\ref{fig:ma}, the green filled circles and blue crosses represent
rocky and icy planets with an insignificant amount of gaseous
envelopes around them.  The green and blue open circles represent
gas-rich rocky and icy planets with gaseous envelopes which are one to
ten times more massive than their cores.  The red filled circles
represent gas giants with gaseous envelopes more than
ten times more massive than their cores.  
The result in panel (a) yields most of gas
giants form with masses an order of magnitude larger than Jupiter.
Even larger masses would be obtainable if the gas accretion is
unimpeded until $M_{\rm p}$ becomes $M_{\rm g,noiso}$ or 
$M_{\rm g,iso} + M_{\rm g,feed}$.  This inconsistency with the observed mass
distribution of the extra solar planets is removed by the thermal gap
formation condition as represented in panels (b) and (c).  However,
the viscous gap formation condition in (d) places an over strict upper
limit on $M_{\rm p}$ so that gas giants cannot form.  Numerical
simulations of disk-planet interaction indicate that $M_{\rm g,th}$ is
the appropriate upper limit for masses of gas giants \citep{Bryden99}.

\subsubsection{Planetary desert}
In panels (a), (b), and (d), these results show a deficit of planets with
intermediate mass (10-100$M_{\oplus}$) and semi major axis less than 3
AU.  The ranges of the deficit is different in panels (a), (b), and (c).  
The lower mass boundary of the planet-deficit domain in
Figure~\ref{fig:ma} is demarcated by $M_{\rm p} \sim 4M_{\oplus}$
for all models.  This value of $M_{\rm p}$ simply corresponds to the
minimum core mass that can satisfy both $M_{\rm c} > M_{\rm c,crit}$ and
$> M_{\rm c,KH}$ (\S 3.1), which are the
condition for the cores to undergo rapid gas accretion.  Transition
across this boundary is weakened by the growth through giant impacts
after the disk depletion.  The truncation conditions of the cores and
gas do not affect this boundary.  Future observational determination
of this lower mass boundary will provide useful information and
constraints on the critical core mass which can lead to the onset of
rapid gas accretion.

The upper mass boundary of the planet-deficit domain in
Figure~\ref{fig:ma} is demarcated by the truncation condition
for gas accretion.  First, we consider the case of $M_{\rm g,iso}$.
The upper mass boundary of the planet-deficit domain corresponds to
the minimum asymptotic mass of gas giants.  As shown in
eq.~(\ref{eq:giants_form_cond1}), in the case of core accretion being
truncated by $M_{\rm c,iso}$, rapid gas accretion would occur if
$f_{\rm disk} \ga f_{\rm tg} \simeq 8 (a/1{\rm AU})^{-1/2}.$ 
In the inviscid-gaseous disk model, the gas accretion
process is terminated by isolation when the planets mass reaches a
value which is given by eq.~(\ref{eq:m_gas_iso0}).  Since $f_{\rm g,0}
= f_{\rm disk}$ in all the calculations described here, we obtain, by
substituting $f_{\rm tg}$ into eq.~(\ref{eq:m_gas_iso0}), a minimum
truncated mass of gas giants in these regions to be
\begin{equation}
M_{\rm g,iso}^* \simeq 1.1 \times 10^3 M_{\oplus}.
\end{equation}
A similar calculation with $f_{\rm tg} \simeq 3 (a/1{\rm
AU})^{-1/2}$ (for cores to be truncated by $M_{\rm c,noiso}$) yields
\begin{equation}
M_{\rm g,iso}^* \simeq 2.5 \times 10^2 M_{\oplus}.
\end{equation}

In regions of the disk where $f_{\rm disk} < f_{\rm tg}$, cores cannot
undergo rapid gas accretion. But, in regions where $f_{\rm disk} > 
f_{\rm tg}$, cores accrete gas until $M_{\rm p}$ reaches $M_{\rm g,iso}^*
\times (f_{\rm disk}/f_{\rm tg})^{3/2}$.  The process of gas accretion
can be stalled by the depletion of the disk gas.  But the time scale
for the planets to attain their asymptotic masses is shorter than the
gas depletion time scale so that only small fraction of planets may
attain some intermediate masses.  Therefore, most planets that undergo
gas accretion have $M_{\rm p} \ga M_{\rm gas,iso}^*$, and $M_{\rm
g,iso}^*$ corresponds to the upper mass boundary of the planet-deficit
domain.  Note that the $a$-dependence vanishes in $M_{\rm g,iso}^*$,
so that the upper mass boundary is essentially independent of $a$ in
panel (a).  If asymptotic mass is regulated by $M_{\rm g,iso}$, the
observed upper boundary of the planet-deficit domain would provide
constraints on the feeding zone width ($\Delta a_{\rm g}$) of gas
giants and the critical surface density of solid components that can
produce cores large enough for gas accretion (equivalently, $f_{\rm tg}$).

\subsubsection{Asymptotic mass of the gas giants}
Next, the results with gas truncation by $M_{\rm g,th}$ are
considered.  These conditions are particularly appropriate since the
ongoing accretion from protostellar disks onto nearly all classical T
Tauri stars clearly reflects the consequence of viscous transport of
angular momentum and mass in the disks.  The thermal truncation
condition does not include an explicit $f_{\rm disk}$ dependence in
the asymptotic mass of the gas giants.  The mass dispersion of giant
planets at given $a$ is caused by the dispersion of the stellar mass
$M_*$ and is relatively small.  The giant planet mass as a function
of $a$ clearly reflects the gap opening condition
(\ref{eq:m_gas_therm}).  The upper mass boundary of the planet-deficit
domain, or equivalently, the lower mass boundary of the giant planets
domain, is demarcated by $M_{\rm g,th}$ with $M_* \sim 0.7M_{\odot}$
in the present models.  The truncation condition for the core growth
stage does not affect the results.  Note that we have adopted some
idealized truncation conditions.  In many numerical simulations, e.g.,
\citep{AL96}, there
are indications that the gas flow on to the giant planets continues,
albeit at some reduced rate, after the gap formation.  Turbulence may
be particularly important in sustaining the planets' growth until they
have acquired larger masses than those shown in panel (b) of
Figure~\ref{fig:ma}.  In order to demonstrate this dependency, we
consider in panel (c) an additional model with $r_{\rm H,c} = 1.5h$.
This relaxed condition yields 3 times larger $M_{\rm p}$ for the gas
giants.

The main difference between panels (b) and (c) is abundance of the green
and blue open circles which denote the gas-rich terrestrial (rocky) planets 
and ice giants
with the mass of the envelope do not exceed ten times that of the core
({\it i.e.} $M_{\rm p} < 10 M_{\rm c}$).  From eqs.~(\ref{eq:m_iso0})
and (\ref{eq:m_gas_therm}), we find
\begin{equation}
M_{\rm g, th} \simeq 0.75 \times 10^3 f_{\rm disk}^{-3/2}
\eta_{\rm ice} ^{-3/2} \left({r_{\rm H,c} \over h} \right)^3
\left({M_\ast \over M_\odot} \right) M_{\rm c, iso}. 
\end{equation}
Although $\eta_{\rm ice}=1$ interior to the ice line, the critical
value of $f_{\rm disk} (\simeq 17)$ for $M_{\rm g, th} < 10 M_{\rm c,
iso}$ in panel b (where $r_{\rm H,c} = h$) is within the range of
$f_{\rm disk}$ (0.1-30) we have adopted in our simulations.  Outside
the ice boundary where $\eta_{\rm ice}=4$, it is much more likely that for
$M_{\rm g, th} < 10 M_{\rm c, iso}$.  Even though the mass of the
gas-rich terrestrial planets and ice giants are dominated by their gaseous
envelopes, the metallicity of their envelope may be 
greater than that of the disk.  But, in panel c where $r_{\rm H,c} =
1.5 h$, $M_{\rm g, th}$ is much larger than $M_{\rm c, iso}$.  The
asymptotic composition of the gas giants in these disks is diluted
by the massive gaseous envelope.

\subsubsection{Radial extent of planet forming region}

The present scenario also indicates an outer boundary for the domain of
gas or ice giant planet formation.  The outer boundary for gas giants
is set by $a_{\rm gi}$ (eq.~[\ref{eq:agi}]) 
where $3 \tau_{\rm c,acc} (M_{\rm c,KH}) = \tau_{\rm disk}$. 
Although in eq.~(\ref{eq:agi}), $a_{\rm gi}$ increases with
$f_{\rm disk}$, it is limited to 30 AU even for the maximum value of
$f_{\rm disk} (=30)$ we have adopted.  This limited extent of the
gas-giant formation zone only applies to the core-accretion scenario.
In the gravitational instability scenario, gas giants are
preferentially formed far from their host stars.
The spatial distribution of gas giants may provide
a discriminating test for the avenue of planet formation.

After the gas depletion, the residual rocky and icy embryos can grow
through giant impacts.  In the above discussions, we have already
indicated that core grow is limited by a) the availability of
building-block material (through $M_{\rm c,noiso}$; eq.~[\ref{eq:no_m_iso0}]) 
and b) the retention efficiency (through $M_{\rm e, sca}$; 
eq.~[\ref{eq:msca}]).  But, even with unimpeded growth and
infinite supply of planetesimals, c) the maximum mass can be attained
within the life span of the host star ($\tau_{\ast}$), through
collisional coagulation, is (eq.~[\ref{eq:m_finalgrow}])
\begin{equation}
M_{\rm e, life} \equiv M_{\rm e} (\tau_\ast) \simeq 0.8 \times 10^4
\left({\tau_\ast \over 1{\rm Gyr}} \right)^3 \left( {\eta_{\rm ice}
\over 4} \right)^3 \left({f_{\rm disk} \over 30} \right)^3  
\left( {a \over {\rm 10 AU} } \right)^{-9} M_\oplus.
\label{eq:mlife}
\end{equation}
The boundary between conditions a) and b) is demarcated 
by $M_{\rm c,noiso} = M_{\rm e, sca}$ and it occurs at
\begin{equation}
a_{\rm ab} = 3 \left( {\eta_{\rm ice} \over 4} \right)^{-1/2} 
\left( {f_{\rm disk} \over 30} \right)^{-1/2} {\rm AU}.
\end{equation}
The boundary between conditions b) and c) is demarcated by 
$M_{\rm e,sca} = M_{\rm c, life}$ and it occurs at
\begin{equation}
a_{\rm bc} = 24 \left({\tau_\ast \over 1 {\rm Gy}} \right)^{2/5}
\left({\eta_{\rm ice} \over 4} \right)^{2/5} 
\left( {f_{\rm disk} \over 30} \right)^{2/5}
{\rm AU}.
\end{equation} 

$M_{\rm e,lfe}$ with $t=1$G years, 
$M_{\rm c,noiso}$, and $M_{\rm e,sca}$ are 
plotted as functions of $a$ for $f_{\rm disk} = 1, 3, 10,$ and 30.
in Fig.~\ref{fig:ice_giants}. 
Rocky and icy planets can exit in the range surrounded by these masses. 
These analytic approximation is consistent with the upper limits
of $M_{\rm p}$ shown in Fig.~\ref{fig:ma}.  In all panels of
Fig.~\ref{fig:ma}, the
formation of both gas and ice giants is quenched beyond $\sim 30-40$
AU despite the availability of planetesimal building blocks.  In
panels a), b), and c), the gas-poor ice giants (represented by the
blue crosses) do not extend above $\sim 20 M_\oplus$.  Their
growth is limited mostly by $M_{\rm e, life}$.  
After the gas depletion, the enhanced velocity dispersion 
makes the accretion rate very low.
In outer regions beyond $\sim 30-40$AU, this effect quenches 
ice giants' further growth.
If the asymptotic mass of the giant planets is limited by the viscous
condition (panel d), gas accretion beyond 10 AU would be quenched when
$M_{\rm p} \sim 40-100 M_\oplus$.  Since their growth is speeded up by
the gas accretion, they may be able to attain masses larger than
$M_{\rm g, vis}$ over some intermediate range of $a$ (3-10 AU) through
collisional coagulation after the gas depletion.

For modest values of $f_{\rm disk}$ such as that in the minimum mass
nebula model ($\sim 1$), $a_{\rm tg} < a_{\rm gi} < a_{\rm ab} <
a_{\rm bc}$.  In this case, rocky, gas and ice giant planets naturally
form in order of increasing distance from their host stars.  This
expectation is consistent with the distribution of the planets in the
Solar System.  Outside $a_{\rm ab}$, the massive ice giants become the
scattering agents which induce the residual planetesimals to be
ejected to form cometary clouds and freely floating planets.  This
extrapolation is also consistent with the observed population of
scattered Kuiper Belt objects \citep{DuncanLevison}.  
Beyond $a_{\rm bc}$, the growth time
is the limiting factor for the emergence of any sizeable protoplanets.
In order to account for the absence of modest-size Kuiper Belt objects
beyond $\sim 40-50$AU, $f_{\rm disk} \la$ a few is required. 

\subsection{Comparison with observational data}

The present observational data (Figures~\ref{fig:obs}) are obtained
with some selection effects.  For example, only planets with radial
velocity amplitude $v_r \ga 10$ms$^{-1}$ and semi major axes $a \la 4-5$AU
have been detected so far.  The planets' mass inferred from these
observations represent a set of minimum values and their real
magnitude is a function of the inclination angle of their orbital
plane to the line of sight.  Many planets are found with $a \sim 0.1-1
$AU and $M_{\rm p} \ga M_{\rm J}$.  Although with the model in which the gas
accretion is truncated by first isolation, we can construct such
massive close-in planets (Figure~\ref{fig:ma}a), but their
emergence requires massive disks which may be prone to gravitational
instability.  The second and third models (Figure~\ref{fig:ma}b,c)
do not produce Jovian-mass planets interior to $\sim 1$AU because the
disks' aspect ratio is relatively small.  There are many possibility,
within the framework of the core accretion scenario, to account for
this discrepancy.  Post-gap-opening accretion can lead to {\it in
situ} formation of such planets.  Orbital migration can also bring in
the relatively massive planets formed in the outer regions of the
disks' aspect ratio is relatively large (see next section).

The present data show an apparent lack of planets with 
$M_{\rm p} \la 200 M_{\oplus}$ and $a \sim 0.2-3$ AU 
(see Fig.~\ref{fig:obs}).  When
corrected for the inclination of their orbits, the demarcation
boundary may be shifted to higher masses, by a factor of two, from that
indicated in Figure~\ref{fig:obs}.  If this hint of a planet-deficit
domain is confirmed by the future observations, it would provide
useful constraints on the truncation condition for gas accretion.  At
their face value, the present data are consistent with the simulated
results with a combination of the isolation condition with $M_{\rm
gas,iso}$ and the thermal condition with $M_{\rm g,th}$.  The
viscous condition with $M_{\rm g,vis}$ limits planet masses to
values much smaller than observed masses unless 1) an extremely large
value of $\alpha$ is taken (which can be ruled out by the persistent
time scale of protostellar disks) or 2) accretion continues after gap
formation.

Our numerical simulations also show an outer boundary for the
planet-deficit domain at around 3 AU.  This boundary reflects the
changes in the dominant process which limits the formation of gas
giants.  At $a \la 3$AU, the condition of $M_{\rm c} \ga 4M_{\oplus}$
is more important, because cores grow rapidly and reach their
asymptotic mass prior to the era of gas depletion.  But, at $a \ga
3$AU, the cores' growth is so slow, that it is more important for
$\tau_{\rm c,acc} \la \tau_{\rm disk}$ to be satisfied.  For most
cores that can start gas accretion, $\tau_{\rm c,acc}$ is of oder of
$\tau_{\rm disk}$.  Hence, many cores may not be able to initiate the
process of gas accretion even though their mass can become larger than
$10 M_{\oplus}$ after gas depletion. As a result, planetary mass
distribution at $a \ga 3$ AU is likely to be continuous and does not
show a clear deficit.  Thus, the outer $a$ boundary of the
planet-deficit domain indicates the critical semi major axis where
core accretion timescales and disk gas depletion timescales are
comparable.

\section{Orbital migration}

The existing data of known extra solar planets include a population of
short-period planets with period down to 3 days.  We consider the
possibility of orbital migration in this section.

\subsection{Model of orbital migration}

There are several suggestions that, due to a torque imbalance,
embedded protoplanetary cores can undergo rapid type I migration
, e.g., \citep{GT80,Ward86}.  At 1 AU, the migration
timescale for $M_{\rm c} \sim 1 M_{\oplus}$ is only $10^4-10^5$years
\citep{Ward97, Tanaka02, Bate03}.  If type I migration is considered,
all the protoplanetary cores have a tendency to rapidly migrate to the
proximity of their host stars, prior to gas accretion.  Thus, type I
migration appears to be inconsistent with prolific formation of extra
solar giant planets as well as the efficient preservation of Earth
mass planets in the habitable zones around the Sun.  

In contrast, type II migration naturally occurs when a planet acquires
an adequate mass to open up a gap.  Gap formation process itself
requires the dissipation of tidally induced density waves.  Turbulent
viscosity is a possible mechanism to provide such a dissipation
mechanism.  It can also regulate the rate of gas accretion onto the
classical T Tauri stars.  Once a protoplanet has acquired an
adequate mass to open up a gap, its orbital evolution is coupled with
the viscous evolution of the disk \citep{LP85}.  

In the next set of models, we include the effect of type II migration
for gas giants which have already attained their asymptotic masses.
For computational convenience, we introduce a simple prescription to
follow the orbital evolution of the gas giant planets. The direction
of the migration can be both inward and outward.  We arbitrarily set
an stopping semi major axis for the inwardly migrating planets to be
$\sim 0.04$AU.  There are several potential mechanisms for halting the
migration near the stellar surface \citep{Lin96, Trilling98}, but none
of these processes are effective beyond $\sim 0.1-0.2$AU.  The
migration can be halted beyond $\sim 0.1-0.2$AU by gas depletion
\citep{Trilling02, Armitage02, Johnstone03}.  However, both the planet
formation and migration timescales at 0.2-3 AU is shorter than
$\tau_{\rm disk} \sim 10^6-10^7$ years.  We consider below the
necessary condition to build up gas giants and deplete the disk during
the course of their migration so that they would be relocated to this
intermediate location.
 
Here we consider only the type II migration in which orbital evolution
of a planet embedded in a gap is coupled with the evolution of the
disk \citep{Lin96, Trilling02, Armitage02}.  In a slowly evolving
(quasi steady) disk, the angular momentum flux due to the viscous
stress is approximately independent of the disk radius.  For a simple
approximation, we estimate the net angular momentum transfer rate, in
the absence of any planets, at the radius of maximum viscous couple
$R_{\rm m}$, e.g., \citep{Lynden-Bell74},
\begin{equation}
\dot{J}_{\rm m} \simeq \frac{3}{2} \Sigma_{\rm g} \nu 
\Omega_{\rm K,m} R_{\rm m}^2 \simeq \frac{3 \alpha}{2} 
\Sigma_{\rm g,m} R_{\rm m}^2
\Omega_{\rm K,m}^2 h_{\rm m}^2,
\end{equation}
where the subscript ``$_{\rm m}$'' denotes values of various
quantities at $R_{\rm m}$.

Across the gap, the planets' tidal torque regulates the angular
momentum transport.  When the disk interior to their orbits is
accreted onto their host stars or that exterior to their orbits 
diffuses to much larger radii, the embedded planets maintain the
angular momentum transport balance $(1/2)M_{\rm p} \Omega_{\rm K,p}
a_{\rm p} \dot{a}_{\rm p} \sim \dot{J}_{\rm m}$ by adjusting their
orbital radius $a_{\rm p}$ \citep{LP85, Lin96}.  This process leads to
planetary migration on a rate,
\begin{equation}
\frac{ \dot{a}_{\rm p} }{a_{\rm p}} \simeq 3 {\rm sign}(a_{\rm p} - R_{\rm m})
\alpha \frac{\Sigma_{\rm g,m}R_{\rm m}^2}{M_{\rm p}}
\frac{\Omega_{\rm K,m}}{\Omega_{\rm K,p}}
\left(\frac{h_{\rm m}}{a_{\rm p}} \right)^2
\Omega_{\rm K,m},
\end{equation} 
where $\Omega_{\rm K,p}$ is Kepler frequency of the planet's orbit at
$a_{\rm p}$.

The sign implies that after it has opened up a gap, an embedded
planet at radius $a_{\rm p}$ would migrate with the disk gas toward
its host star if $a_{\rm p} < R_{\rm m}$ and away from the star if
$a_{\rm p} > R_{\rm m}$.
Using eqs.~(\ref{eq:h_a}) and (\ref{eq:sigma_g}) to substitute for
$\Sigma_{\rm g}$ and $h$, we obtain a migration timescale,
\begin{equation}
\tau_{\rm mig} =  \frac{a_{\rm p}}{\mid \dot{a}_{\rm p} \mid} =
0.8 \times 10^6 f_{\rm g}^{-1}
\left(\frac{M_{\rm p}}{M_{\rm J}}\right)
\left(\frac{M_{\odot}}{M_*}\right)
\left( \frac{\alpha}{10^{-4}} \right)^{-1}
\left( \frac{a_{\rm p}}{1{\rm AU}} \right)^{1/2} \, {\rm years}.
\label{eq:tau_mig}
\end{equation}

\subsection{An $\alpha$-prescription for gas disk evolution}

The value of $R_{\rm m}$ depends on the $\Sigma_{\rm g}$
distribution.  For the minimum-mass disk model, we estimate it to be
$\sim 10$AU \citep{LP85}.  However, the observationally inferred size
and mass of the disks have a large dispersion.  A similar dispersion
are expected in $R_{\rm m}$.  During the viscous diffusion process,
$R_{\rm m}$ evolves outward if there is no infall to replenish the
disk mass.  In the absence of infall and outflow, $\Sigma_{\rm g}$
declines in most part of the disk as mass is accreted onto the host
stars.  From the conservation of the total disk mass
\begin{equation}
J_{\rm tot} \sim \pi \Sigma_{\rm g} R_{\rm m}^4 \Omega_{\rm K,m},
\end{equation} 
we expect 
\begin{equation}
\frac{\dot{R}_{\rm m}}{R_{\rm m}} \sim -
\frac{2\dot{\Sigma}_{\rm g}}{5\Sigma_{\rm g}}.
\label{eq:R_m0}
\end{equation}
If the disk depletion is mostly driven by photo evaporation or stellar
wind ablation, the expansion rate of $R_{\rm m}$ would be slower.
However, as shown below, in the model with $\Sigma_{\rm g} \propto
a^{-3/2}$ and $h \propto a^{-5/4}$ (equivalently, $T \propto
a^{-1/2}$), planetary migration rate is independent of $R_{\rm m}$
except for outer regions ($a \ga R_{\rm m}$) where the direction of
migration is outward.  Hence, we only need an approximate prescription
for the time evolution of $R_{\rm m}$ in our simulation.
Corresponding to eqs.~(\ref{eq:gas_decay}) and (\ref{eq:R_m0}), we set
\begin{equation}
R_{\rm m} = 10 \exp\left(\frac{2t}{5 \tau_{\rm disk}}\right) \, {\rm AU}.
\label{eq:R_m}
\end{equation}

In a quasi steady state, the radial velocity of the gas which is
\begin{equation}
u_r \simeq - \frac{3}{2} \frac{\nu}{a_{\rm p}} \simeq
- \frac{3 \alpha}{2} \left(\frac{h_{\rm p}}{a} \right)^2 a_{\rm p}
 \Omega_{\rm K,p},
\end{equation}
in regions well inside ${R_{\rm m}}$. The scale $h_{\rm p}$ is
evaluated at $a_{\rm p}$.  Note that $\vert \dot{a}_{\rm p} \vert$
cannot be faster than $\vert u_r \vert$.  Equivalently,
$\tau_{\rm mig}$ cannot be smaller than
\begin{equation}
\tau_{\rm disk,acc} = \frac{a_{\rm p}}{\mid u_r \mid} \simeq 
4.3 \times 10^5 
\left( \frac{\alpha}{10^{-4}} \right)^{-1}
\left( \frac{a_{\rm p}}{1{\rm AU}} \right) \, {\rm years}.
\label{eq:disk_acc}
\end{equation}
For small planets embedded in relatively massive disks,
$\tau_{\rm mig}$ is limited by $\tau_{\rm disk,acc}$.  Also, when
$a_{\rm p}$ decreases to 0.04AU, migration is halted as mentioned
above.

Integrating $\dot{a}_{\rm p}$ with eqs.~(\ref{eq:tau_mig}) and
(\ref{eq:R_m}), we show examples of type II migration with our model
in Figure~\ref{fig:t_mig}, where time is scaled by $\tau_{\rm mig}$ at
1AU ($\tau_{\rm mig,1}$).  At $a \sim 1$AU, planets can migrate by a
significant distance in the limit that $\tau_{\rm dep} \ga \tau_{\rm
mig,1}$.  Since $\tau_{\rm mig} \propto a_{\rm p}^{1/2}$, the inner
planets tend to migrate over a greater radial extent and the inward
migration accelerates with time.  In all figures, we indicate the
evolution of $R_{\rm m}$ with dashed lines.  Planets exterior to
$R_{\rm m}$ migrated outward.  Some planets which formed slightly
outside the initial $R_{\rm m}$ migrate outward first.  But, as
$R_{\rm m}$ expands beyond their orbits, their migration reverses
direction and they undergo orbital decay \citep{LP85}.  The migration
of some planets are stalled at intermediate radii when the disk is
depleted.  Note that the number of planets which stalled with
intermediate semi major axes or equivalently periods is a function of
$\tau_{\rm dep} /\tau_{\rm mig, 1}$.  The initial migration time scale
at 1AU, $\tau_{\rm mig, 1}$, provides an arbitrary fiducial comparison
between the disk depletion and planets' migration time scales.  In
models with $\tau_{\rm dep} /\tau_{\rm mig, 1} > 1$, delicate timing
is needed for a few planets to stall their orbital evolution at a
fraction of an AU.  This result is similar to those found by
\citet{Trilling02} and \citet{Armitage02}.  
But, in models with relatively small
$\tau_{\rm dep} /\tau_{\rm mig, 1}$, a significant fraction of the gas
giant planets formed near 1 AU may attain asymptotic semi major axes
between 0.04 to 1 AU.

In Figure~\ref{fig:t_mig}, we highlight the critical migration paths
of (1) those planets which marginally migrate from an initial semi
major axis $a_1$ to 0.04AU and (2) those planets which attain an
asymptotic semi major axis which are greater than 0.9 times that of
their initial semi major axis $a_2$. 
Since planets form with semi
major axis between $a_1$ and $a_2$ attain asymptotic $a$'s between
0.04 AU and $\sim a_2$, relatively large values of $\tau_{\rm dep}
/\tau_{\rm mig, 1}$ would lead to a significant modification in the
semi major axis distribution of the planets.

\subsection{The modification of the semi major axis distribution by migration}

We apply the above prescription for migration to the planetary growth
model.
We added the effects of migration to the Monte Carlo calculations in
\S 4.2 (Figures~\ref{fig:ma} a, b, and c).  
We assume that type II migration starts when a planet's mass
reaches $M_{\rm g, th}$ and a gap is opened up in the nascent disk
near its orbit (models (a) and (b)).  
In the case that the gas truncation is induced by
isolation, it is less clear when type II migration would be initiated.
For simplicity, we assume that migration starts when planetary mass
reaches $M_{\rm g, iso}$ which is larger than $M_{\rm g, th}$.
For model (c), migration is
initiated when $r_{\rm H}$ becomes larger than $h/1.5$
such that there is a range for which gas accretion and migration occur
concurrently.  
This migration condition is similar to $M_{\rm p} > M_{\rm g,vis}$.
(As mentioned in the last section, the gas truncation by $M_{\rm gas, vis}$
seems to be inconsistent with the observational data, but
the migration condition by $M_{\rm gas, vis}$ may be reasonable.)

In these calculations, $\tau_{\rm dep} = 10^6-10^7$
years.  The time dependent calculation of disk evolution \citep{Lynden-Bell74}
indicates that the disk mass declines on the viscous diffusion time
scale near $R_{\rm m}$.  If gas depletion in disks is due to their
viscous evolution, we would expect $\tau_{\rm dep}$ to be comparable
to $\tau_{\rm disk,acc}$ (eq.~[\ref{eq:disk_acc}]) near $R_{\rm m} \sim
10$AU.  In order to match the observed properties of protostellar
disks around classical T Tauri stars, we adopt $\alpha = 10^{-4}$
which corresponds to $\tau_{\rm dep} / \tau_{\rm disk,acc} \sim 1$ at
10AU.

The results of our simulations are shown in Figs.~\ref{fig:ma_mig} for
three series of models.  In each case, the gas and core accretion are
truncated by the conditions which correspond to those in
Fig.~\ref{fig:ma}, respectively.  
The results show that the spatial
distribution of the gas-poor cores is not affected by the migration
because it only affects those planets which are able to accrete gas
and to open up gaps.  But for gas giant planets, equation
(\ref{eq:tau_mig}) indicates that the migration time scale increases
with their masses and semi major axes. {\it The less massive gas
giants are formed preferentially with relatively small semi major axes
and they migrate to $\sim 0.04$AU in all the cases.}  This result is
consistent with the observed mass distribution of the short-period
planets which appears to be smaller than that of planets with periods
longer than a few months \citep{Udry03}.

Gas giant planets with $\tau_{\rm mig} \la \tau_{\rm disk}$ migrate
over extended radial distance provided the disk gas is preserved for a
sufficiently long time for them to form.  For example, the critical
value of $f_{\rm disk}$ for the formation of gas giants is $\sim 3-8$ at
$a \sim 1$AU where $\eta_{\rm ice} =1$ (see \S \ref{sec:sec41}).  
From eq.~(\ref {eq:pl_acc_rate}), we find that in disks with
$f_{\rm disk}$ larger than the critical value, the time scale for a core to
acquire a mass $M_{\rm p} \ga 4 M_\oplus$
is $3 \tau_{\rm c, acc} \la 10^5$yrs which is much smaller than the mean
disk depletion time scale $\tau_{\rm disk} \sim 10^6-10^7$ years.
If the subsequent gas accretion is limited by either $M_{\rm g, th}$
or $M_{\rm g, iso}$, the asymptotic mass of the gas giant would be
$\la 10 M_{\rm J}$ (Figs.~\ref{fig:ma}).  
For this range of
$M_{\rm p}$ and $f_{\rm disk}$, we find, from eqs.~(\ref{eq:tau_mig}) and
(\ref{eq:disk_acc}), that the migration time scale $\tau_{\rm mig}$ is
less than a few Myr provided $\alpha \sim 10^{-4}$.  In most disks, 
$R_{\rm m} \gg 1$AU
so that the planets formed at $a \sim 1$AU have $\tau_{\rm c, acc} \la
\tau_{\rm mig} \la \tau_{\rm disk}$.  Thus, within our model here, 
{\it most gas giant planets
formed within $\sim 1$AU are removed by the orbital migration
independent of the truncation condition for gas accretion.}

Nevertheless, there is a population of gas giants which attain
asymptotic semi major axis in the range of a fraction of an AU.  In
fact, the distribution of semi major axis is nearly continuous from
$\sim 0.04-4$AU.  As we have indicated in the discussion of
Figure~\ref{fig:t_mig}, planets formed between $a_1$ and $a_2$
slightly outside the initial radius of maximum viscous stress ($R_{\rm
m}$) migrate outward first.  They reverse their direction and migrate
inward after $R_{\rm m}$ has expanded beyond their orbit.  (In all our
calculations, we adopt $R_{\rm m} =10$AU initially so that $a_1 \sim
3-5$AU and $a_2 \sim 10$AU.)  Their period distribution
in the range of a few weeks to years is determined by the ratio of the
disk depletion and viscous time scales. 
With our adopted value
$\tau_{\rm dep} / \tau_{\rm disk, acc} (\simeq 1)$, the simulated $a$
distribution is similar to that observed.

Numerical simulations of disk-planet interaction suggest that there
may be lingering gas accretion after the gap formation which may lead
to $M_{\rm p}$ to be somewhat but not substantially larger than
$M_{\rm g,th}$.  The post formation accretion
may be particularly effective for the eccentric planets because their
large epicyclic excursion can be extended beyond the boundaries of the
gap.  From eq.~(\ref{eq:tau_mig}) we find that $\tau_{\rm mig} \simeq
\tau_{\rm disk}$ when $M_{\rm p}$ reaches 
\begin{equation}
M_{\rm mig} = 12.5 f_{\rm g} \left( {\tau_{\rm disk} \over 10^7
{\rm years} } \right) \left( { M_\ast \over M_\odot} \right) 
\left( { \alpha \over 10^{-4} } \right) \left({ a \over 1
{\rm AU}} \right)^{-1/2} M_{\rm J}.
\end{equation}
If such an effect leads to the formation of any massive
planets with $M_{\rm p} >M_{\rm mig}$, they would not be able to
migrate significantly from their birth place.

\subsection{Preservation of the planet-deficit domain}

The results in Figures~\ref{fig:ma_mig} show that the domain of the
planet desert becomes more clearly defined with all three prescriptions for
the truncation of gas accretion.  In comparison with the planets'
initial properties (see Figs.~\ref{fig:ma}), {\it the effect of the
orbital migration is to reduce and erase the differences introduced to
the $M_{\rm p}-a$ distribution by the different truncation conditions
for gas accretion.}

In this series of simulations, we only consider type II migration,
which occurs only if a planet can open up a gap with a mass in excess
of (0.3-1)$M_{\rm g,th}$ 
[$\sim (30-10^2) \times (a/1{\rm AU})^{3/4} M_{\oplus}$].  The
lower-$M_{\rm p}$ boundary of the planet desert is not modified by the
type II migration process because it is populated by cores which have
not yet initiated the process of rapid gas accretion and may not
migrate.  As we have already shown that this boundary is primarily
determined by the cores' mass required for transition from the
predominantly planetesimal to the mostly gas accretion, it is
insensitive to the truncation condition for the cores and their
asymptotic masses.

A direct comparison between Figures~\ref{fig:ma} and
\ref{fig:ma_mig} clearly indicates that the planet migration process
significantly alter the upper-$M_{\rm p}$ boundary of the
planet-deficit domain.  Gas giants formed with $a < a_1 \sim 3-5$AU
(see Figs.~\ref{fig:ma_mig}) migrate to 0.04 AU.  But some planets
formed between $a_1$ and $a_2 \sim 10$AU, attain asymptotic $a$
between $0.04-4$AU.  As the semi major axis of these marginally
survivable gas giants spread out, they attain a nearly logarithmic
period distribution similar to the observed period distribution.
Outside $a_2$, the gas giants form in the advanced stages of disk
evolution when there is insufficient gas to induce any significant
migration.  Based on the results in Figures~\ref{fig:ma_mig}, we
infer a substantial upturn in the semi major axis 
distribution of Jupiter-mass
gas giants at around 10AU. The boundary of the upturn in
the period distribution is determined by the ratio of gas depletion 
and the viscous evolution time scales of the disk.

During the migration, the gas giants are not expected to lose mass.
Thus, the upper-$M_{\rm p}$ boundary of the planet-deficit region
is set by the lowest gas-accretion-truncation mass between $a_1$ and
$a_2$.  Above this upper boundary, the spread in $M_{\rm p}$ among the
gas giants with any given $a$ is due to the dispersion in the
gas-accretion-truncation condition induced by the mass range of the
host stars and nascent disks.

In this series of simulations, the outer-$a$ boundary of the planet
desert in the $M_{\rm p}-a$ diagram occurs at $\sim 3$ AU for all
three classes of models.  {\it The magnitude of the outer-$a$ boundary
is determined by the location where cores can attain a sufficient mass
to undergo efficient gas accretion prior to the depletion of the gas
in the disk.}  Since this condition is associated with the evolution
of the cores, the location of the outer-$a$ boundary does not depend
on either the gas-accretion-truncation conditions or the migration
process.

Beyond the outer-$a$ boundary, ice-giant planets, similar to Uranus
and Neptune, can form with $10-100M_{\oplus}$.  In these region, the
cores grow relatively slowly.  Even though the asymptotic mass of the
cores is relatively large, they are unable to acquire $M_{\rm c,KH} \sim 4
M_\oplus$ to initiate a phase of rapid gas accretion.  The large
asymptotic mass of these planets is acquired during the final phase of
their assemblage after gas in the disk has been depleted.  In the limit
that some residual gas in the disk is preserved for an extended length
of time, type II migration cannot relocate these planets because their
mass is too small to open up a gap.  Nevertheless, the mass spectrum
at any $a$ beyond the outer boundary of the planet desert can evolve
slightly with gas accretion and significantly via residual core
accretion and giant impacts.  The time scale of growth through giant
impacts is similar to that for the oligarchic cores and is an
increasing function of the planets' mass.  This further coagulation
process leads to a continuous mass spectrum.  The range of ice
giants' masses in the region just beyond the outer-$a$ boundary is
determined by the dispersion in the heavy element content, {\it i.e.}
$f_{\rm disk}$, out there.

The center of the planet-deficit domain is not entirely void of
planets with intermediate $M_{\rm p}$ and $a$'s.  Within the present
theoretical framework of our models, only type II migration can modify
the planets' semi major axis.  But the necessary condition for type II
migration is for gas giant to reach their gas-accretion-truncation
mass $M_{\rm g, th}$ which is an increasing function of $a$.  Gas
giants with $M_{\rm p}$ less than the local $M_{\rm g, th}$ can be
formed interior to any disk location.  However, planets formed
interior to the radius of maximum viscous stress in the disk $R_{\rm
m}$ must always migrate inward.  Since $R_{\rm m}$ is generally
comparable to or larger than $a_2$, planets with intermediate $M_{\rm p}$
and $a$ cannot migrate into the planet-desert domains from smaller or
larger $a$'s.  In our simulation, a few planets located in the
planet-deficit, intermediate $M_{\rm p}$ and $a$ region can be formed in
marginally massive disks where the $f_{\rm disk} > f_{\rm disk, min}$
and $f_{\rm g} \sim (M_{\rm p} / M_{\rm J}) (a /1 {\rm AU})^{-1/2}$
(see eqs.~[\ref{eq:fcrit}] and [\ref{eq:m_gas_non_iso}]).  In this
estimate, we assume that since these planets do not have adequate mass
to open a gap, they eventually acquire all the nearby residual disk
gas.

\subsection{The spatial extent of planetary systems}

In the scenario we presented here, we artificially halt the migration
at 0.04 AU to represent the effect of tidal and magnetospheric
interaction between short-period planets and their host stars.
Besides these processes, MHD turbulence in the disk may also play a
role in halting their migration \citep{Terq03}. Since most of the
short-period planets were formed with $a < a_1$, they generally form
with less mass than the planets which accreted gas later, at larger
disk radii, and retained a greater fraction of their original semi
major axis. Most gas giant
planets formed interior to $a_1$ migrate to $0.04$ AU.  Since gas
giants can readily form well within $a_1$ (see Figs.~\ref{fig:ma}),
a large accumulation of short-period planets is expected.  Although
there is a spike in the observed period distribution just outside 0.03AU, the
amplitude of this enhancement is modest.  Near the surface of the host
stars, the combined effects of tidal, magnetospheric, and radiative
effects of the host stars as well as the secular perturbation of more
distant planets, disks, and stars may lead to the destruction of the
short-period planets and the disruptions of their orbits.  

For the outer boundary of planetary systems, we find that there are
several very massive planets which managed to migrate to $a \sim
50-100$AU in panels a), b), and c) of Figure \ref{fig:ma_mig}.  
These planets were formed at 10-30 AU, just outside $a_2$
in massive disks we have simulated.
With their large values of $a \sim 10$-30AU and $f_{\rm disk}\sim 20$-30,
the gravitational stability parameter $Q$ (eq.~[\ref{eq:toomre_q}]) 
at formation locations of these planets are found to be 0.7-1.2,
which may be (marginally) gravitationally unstable.
Hence, the outward migrating massive gas giants may 
actually be more rare.

For disks with modest masses, ice giants can accrete gas to form
sizable gaseous envelopes (represented by open blue circles) well
beyond the ice boundary.  Although we assume that the ice giants do
not migrate because they do not have a sufficient mass to open a
clean gap.  Nevertheless, the weak tidal interaction may be adequate
to induce the ice giants to migrate primarily outward over a limited
extent.  In the Solar System, the migration of Neptune over several AU's
on a time scale of $\sim 10$ Myr may have led to the capture of Kuiper Belt 
objects on its 3:2 (but not on its 2:1) mean motion resonances \citep{Ida00}.

In the current analysis, we do not consider the interaction between
multiple planets which can lead to dynamical instabilities
\citep{Chambers96, LI97, Marzari02}.  Although
sizeable planets may be ejected to large distances from their host
stars, such an instability is likely to induce the orbits of the
residual planets to become highly eccentric and dynamically segregated
with large ($\ga 10$) period ratios, e.g., \citep{Weiden96, LI97}.
The search for the orderly-packed
ice giants with nearly circular orbits may provide useful constraints to
differentiate the core accretion and gravitational instability
scenarios for planet formation.

\section{Summary and Discussions}

We have investigated the mass and semi major distributions of extra
solar planets through theoretical models based on the core accretion
model for formation of gas giants.  Core accretion from planetesimals
is modeled by the results of N-body simulations.  Gas accretion onto
the cores is modeled with Kelvin-Helmholtz contraction of gas
envelope.  Since the termination of gas accretion onto cores have some
uncertainty, we considered various models for the asymptotic mass of
the gas giants.  
In some calculations, we also include the effects of
type II migration.

In general, the formation of gas giants is favored just outside the
ice boundary at $\sim 3$AU.  Inside the ice boundary, because the
volatile gases cannot condense into grains, the mass of solid cores
tend to be smaller than $4 M_{\oplus}$ which is necessary for the
onset of rapid gas accretion onto them.  On the other hand, in outer
regions ($\ga 10$AU), core accretion is so slow that cores cannot
attain $4 M_{\oplus}$ before the disk gas is depleted on the time
scale of $\sim 10^6-10^7$ years.  In very massive disks, however,
cores become massive enough for the onset of rapid gas accretion even
inside the ice boundary within the disk life time to attain gas mass
$\ga 100 M_{\oplus}$.  Since planets grow very rapidly from $10
M_{\oplus}$ to $100 M_{\oplus}$, planets of the intermediate masses
are rare at $a \la 3$AU (``planet desert'').  If the effect of type II
migration is included, the deficit becomes more clear and tends to
become independent of the model of termination of gas accretion.  It
also results in accumulation of planets at $a \sim 0.03-0.1$AU.  As a
result, the deficit becomes $\sim 10-100 M_{\oplus}$ and $0.2-3$AU.

The current radial velocity survey of nearby stars is limited by the
availability of accurate radial velocity measurements (down to the
level of $v_r \ga 10$ ms$^{-1}$) over the past few years.  Planets with
masses comparable to that of Jupiter and $a \la 4$-5AU are inferred
from these observations.  The processed data show a hint of the planet
desert (Figure~\ref{fig:obs}).  Further observations in the regime of
longer periods (larger $a$) and smaller $v_r$ will clarify the
existence of a planet desert (a deficit of planets with intermediate
mass and semi major axis planets).  

Gas giants could also be formed by self-gravitational collapse of a
disk, e.g., \citep{Boss01}.  If extra solar planets are formed from
the self-gravitational collapse, the mass and semi major axis
distributions of extra solar planets should show very different
features from those we show here based on the core accretion model for
gas giants.
As we showed in \S 4 and 5, with the core accretion model, gas giants 
may be limited to the regions $a \la 20$AU, except for planets
on eccentric orbits that are scattered from inner regions
by dynamical instabilities in multiple gas giant systems
(which we do not consider in this paper).
On the other hand, in the self-gravitational collapse model,
gas giants are preferentially formed far from their host stars
(see eq.~[\ref{eq:toomre_q}]), although the mass distribution
of gas giants is not clear.
The search for gas giants on nearly circular orbits beyond
20-30AU may provide useful constraints to
differentiate the core accretion and gravitational instability
models for formation of gas giants.

If the the presence of a planet desert is confirmed by future
observations, the boundaries of the planet-deficit domain in the mass
and semi major axis plane should provide the following constraints on
the planetary formation processes (see also Fig.~\ref{fig:desert}):
\begin{itemize}
\item The lower mass boundary ($M_{\rm l}$) indicates the core mass that can
initiate the onset of rapid gas accretion.  Our theoretical model
predicts that $M_{\rm l} \sim 4M_{\oplus}$ and it does not depend on $a$.
(If we include the coalescence between cores after gas depletion,
$M_{\rm l}$ slightly increases with $a$.)
\item The value of upper mass boundary ($M_{\rm u}$) and its dependence on
$a$ provide clues on the truncation mechanism of gas accretion.  For
various models we find (1) $M_{\rm u} = 3 \times 10^2 - 10^3 M_{\oplus}$
(and independent of $a$) if the gas giants' growth is limited by an
isolation process, (2) $M_{\rm u} = 20 (\alpha/10^{-3}) (a/1{\rm AU})^{1/2}
M_{\oplus}$ if a complete gap is formed through the viscous condition,
or (3) $M_{\rm u} = 80 (a/1{\rm AU})^{3/4} M_{\oplus}$ if it is due to the
thermal condition.  However, if the effect of type II migration is
included, the results in models (1) and (3) lead to similar $M_{\rm
p}-a$ distribution.  Then the magnitude of $M_{\rm u}$ constrains the
migration timescales and disk depletion timescales.
\item The smaller $a$ boundary ($a_{\rm l}$)
indicates the maximum radius at which type II migration can be
halted.
\item The larger $a$ boundary ($a_{\rm u}$) traces out the radius 
where a dominant condition for 
formation of gas giants changes between $\tau_{\rm c,acc}$
condition and $M_{\rm c,iso}$ condition.
At $a \ga a_{\rm u}$, the former is more important than the latter,
and vice versa.
Our theoretical model predicts that $a_{\rm u} \sim 3$ AU.
\end{itemize}

The results presented here show that the sequences of rocky (terrestrial), 
gas giant
and ice giant planets are clearly segregated in the $M_{\rm p}-a$
diagram.  The importance of this diagram to planet formation is
similar to that of the color-magnitude diagram for stellar evolution.
Comparison of future observations and more refined theoretical models
will reveal the formation processes of extra solar planets.  In the
near future our inability to directly detect the presence of
terrestrial planets limits the comparison to only the giant planet
population.  In follow-up papers, we will consider a wider variety of
models and identify some robust constraints.  Our eventual objective
is extrapolate the ubiquity of rocky planets and formulate a
predictable theory of planet formation from the statical properties
of the giant planets.

\acknowledgements 
We thank E. Kokubo, M. Ikoma, and G. Laughlin for useful discussions.
This work is supported by the NASA through NAGS5-11779 under 
its Origins program, JPL 1228184 under its SIM program, 
and NSF through AST-9987417.

\clearpage

\begin{figure}
\plotone{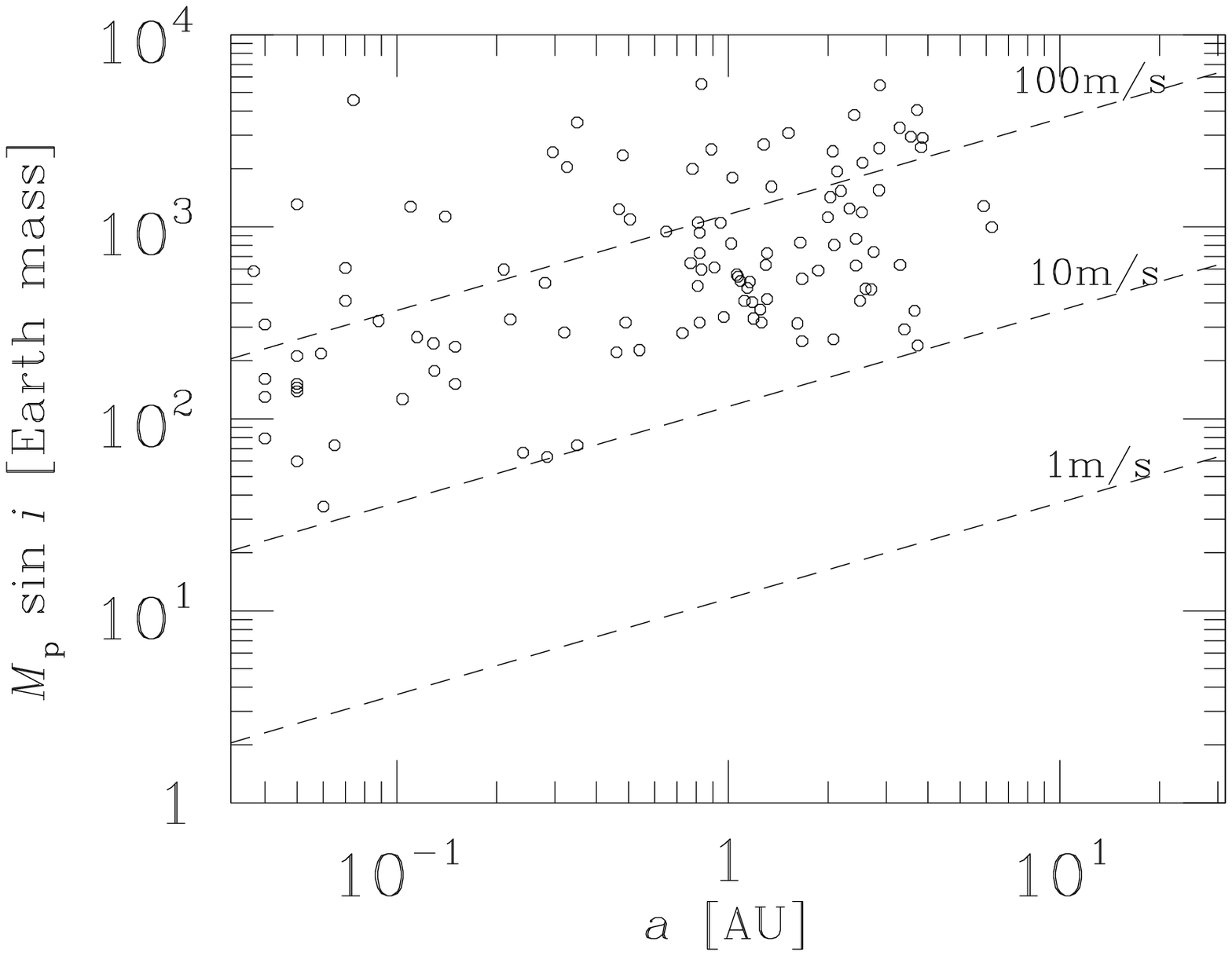}
\caption{
Distributions of semimajor axes ($a$) and masses ($M_{\rm p}$) of
discovered extrasolar planets are shown by open circles.
Unit of mass is Earth mass $M_{\oplus}$
(Jupiter mass is $M_{\rm J} \simeq 320M_{\oplus}$).
$i$ is the angle between an orbital plane and the line of sight. 
The dashed ascending lines correspond to radial velocity amplitude of
100 ms$^{-1}$, 10 ms$^{-1}$, and 1 ms$^{-1}$ (assuming the host star mass is $1M_{\odot}$).
}
\label{fig:obs}
\end{figure}

\begin{figure}
\plotone{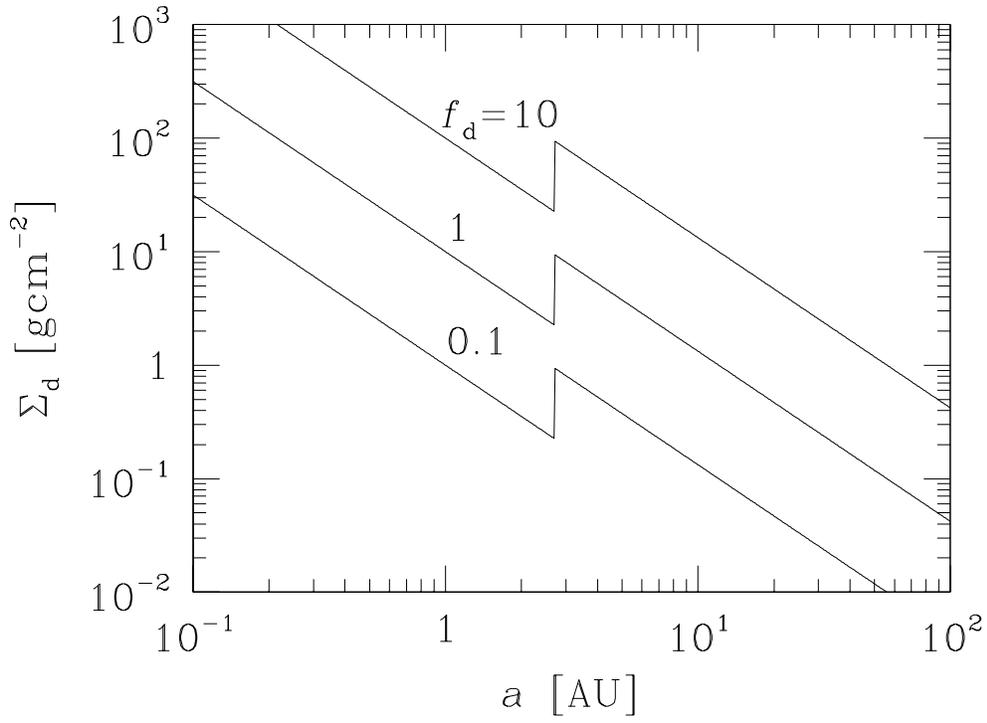}
\caption{
$\Sigma_{\rm d}$ as a function of $a$ that we used.
$f_{\rm d}$ is an enhancement factor from the minimum-mass disk
model for our Solar system.
The jumps at 2.7AU are caused by ice grain condensation
at $a > 2.7$AU.
}
\label{fig:sig_dust_distr}
\end{figure}

\begin{figure}
\plotone{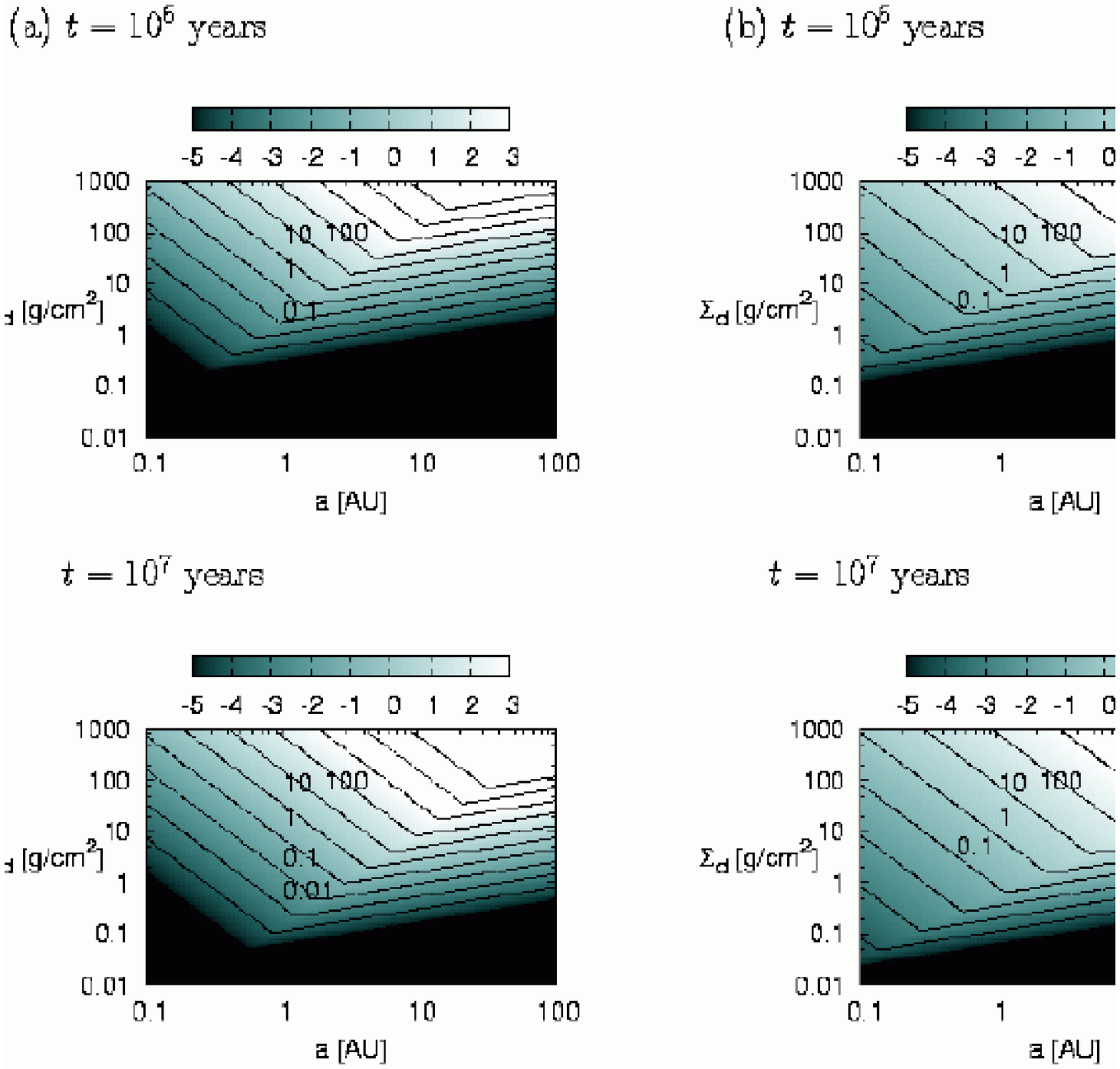}
\caption{
(a) Analytical estimate of the core mass after $10^6$ years
(upper panels)
and $10^7$ years (lower panels).
Core mass is truncated by
$M_{\rm c,iso}$.  Initial core mass is $10^{20}$g.
We used $m=10^{18}$g, $\Delta a_{\rm c} = 10r_{\rm H}$, 
and $\Sigma_{\rm g}/\Sigma_{\rm d}=240$.
Labels in the contours are $M_{\rm c}/M_{\oplus}$.
(Numbers of color box are $\log_{10}(M_{\rm c}/M_{\oplus})$.)
(b) The same plots except for
the truncation by $M_{\rm c,noiso}$ instead
of $M_{\rm c,iso}$. 
}
\label{fig:core_a_sig}
\end{figure}

\begin{figure}
\plotone{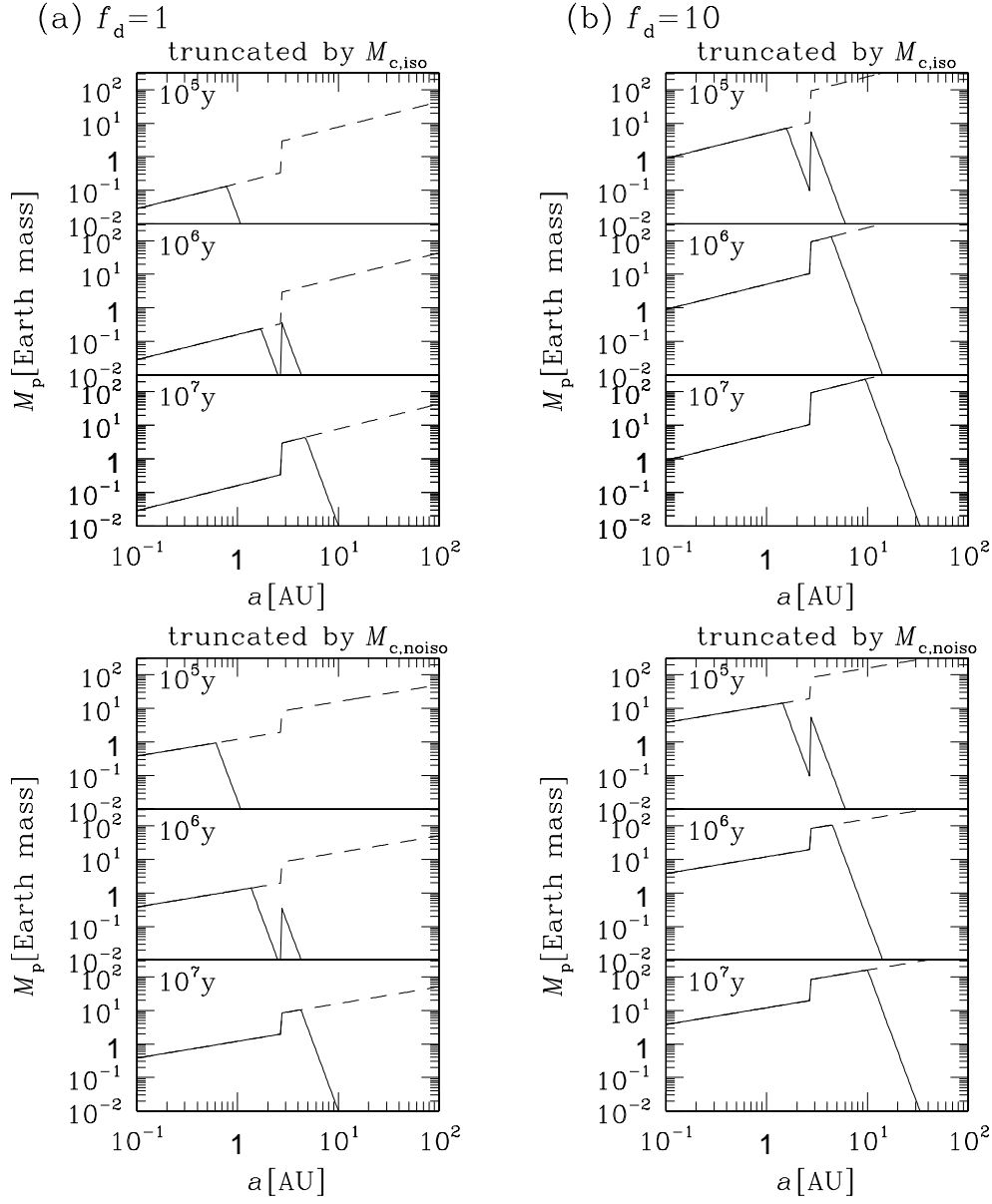}
\caption{
The time evolution of core mass ($10^5$, $10^6$, and $10^7$ years)
is shown by sold lines, in the case of (a) $f_{\rm d} = 1$
and (b) $f_{\rm d} = 10$.  
Core mass is truncated by
$M_{\rm c,iso}$ (upper panels) or by $M_{\rm c,noiso}$
(lower panels), which
are shown by the broken lines.
The jumps at 2.7AU are caused by increase in solid materials
due to ice condensation. 
}
\label{fig:core_evolve}
\end{figure}

\begin{figure}
\plotone{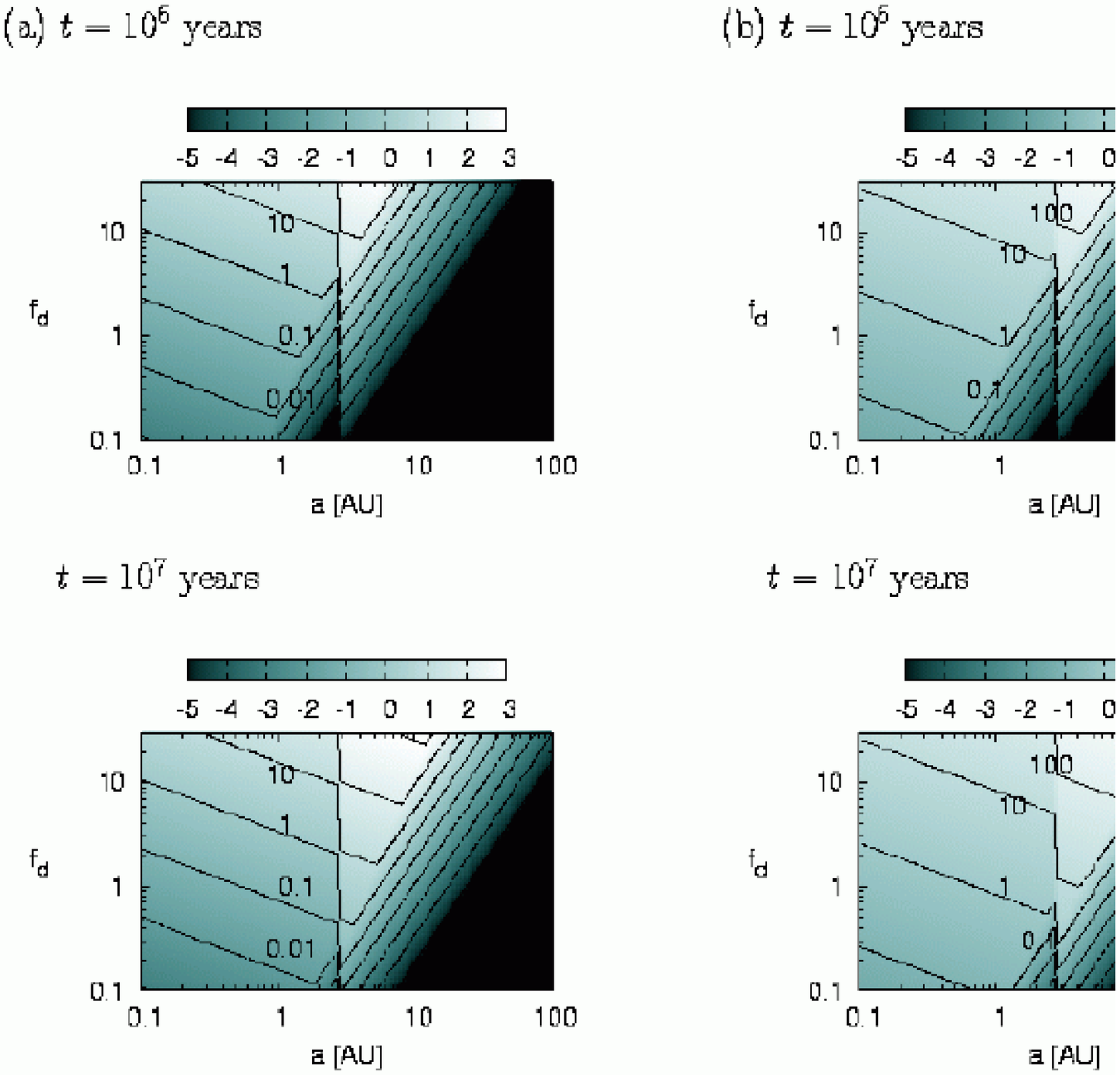}
\caption{
Core masses after $10^6$ years (upper panels) and $10^7$ years 
(lower panels) as a function of $a$ and
$f_{\rm d}$.
(a) Core mass is truncated by $M_{\rm c,iso}$;
(b) the truncation by $M_{\rm c,noiso}$.
}
\label{fig:core_a_fdust}
\end{figure}

\begin{figure}
\plotone{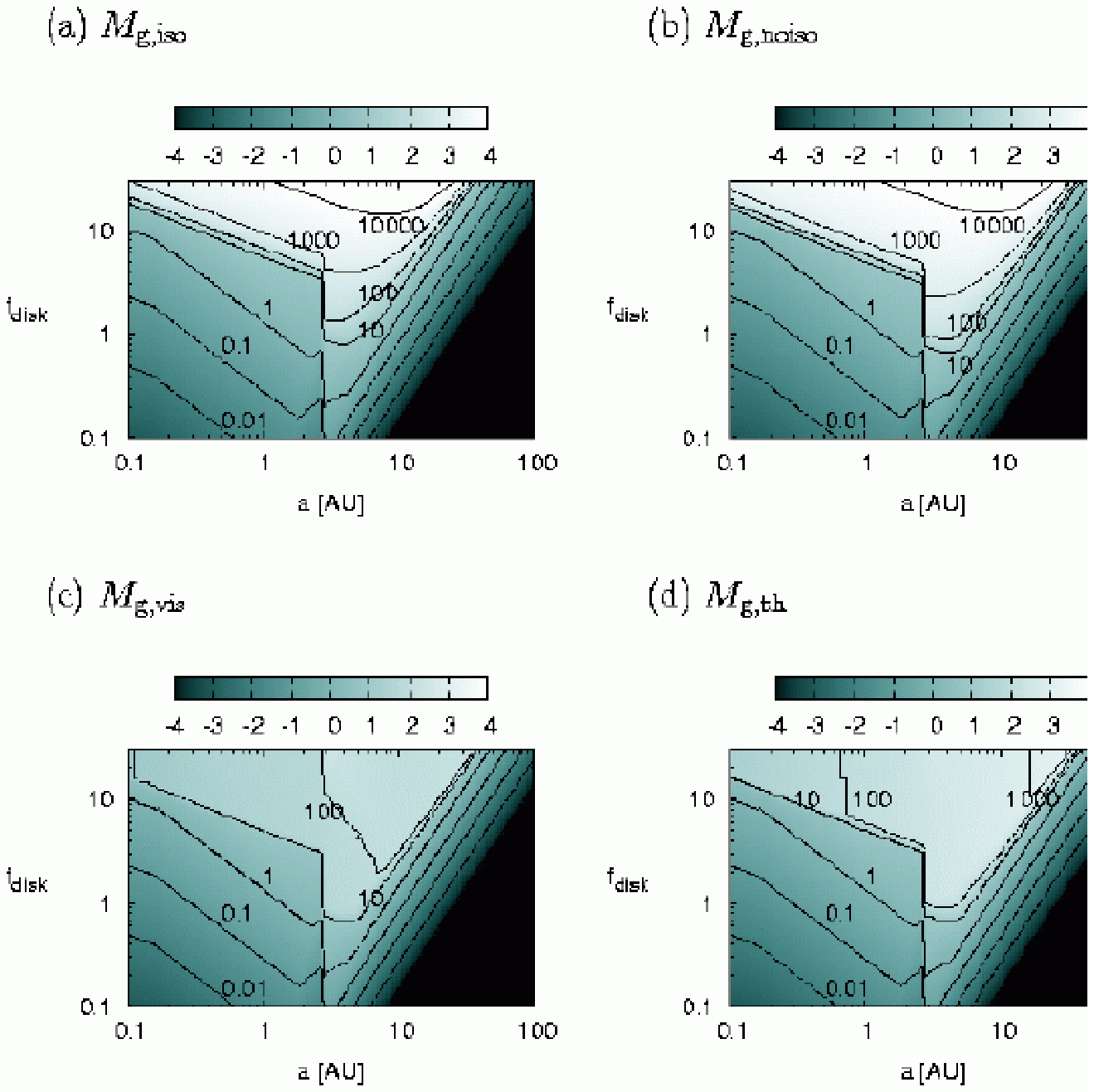}
\caption{
Planet masses including gas envelope at $10^9$ years as a function of
$f_{\rm disk}$ ($=f_{\rm d}=f_{\rm g,0}$) and $a$.
$\tau_{\rm disk} = 10^7$ years and $a_{\rm ice} = 2.7$AU are adopted.
Cores are truncated by $M_{\rm c,iso}$ (after $f_{\rm g}$ declines below
$10^{-3}$, truncated by $M_{\rm e,iso}$).  Gas accretion
is truncated at (a) $M_{\rm g,iso}$, (b) $M_{\rm g,noiso}$, 
(c) $M_{\rm g,vis}$, and (d) $M_{\rm g,th}$.
We adopt $\Delta a_{\rm g} = 2r_{\rm H}$ in (a),
$M_* = 1 M_{\odot}$ in (c) and (d), and
$\alpha = 10^{-3}$ in (c).
Labels in the contours are $M_{\rm p}/M_{\oplus}$.
(Numbers of color box are $\log_{10}(M_{\rm p}/M_{\oplus})$.)
}
\label{fig:core_gas_iso}
\end{figure}

\begin{figure}
\plotone{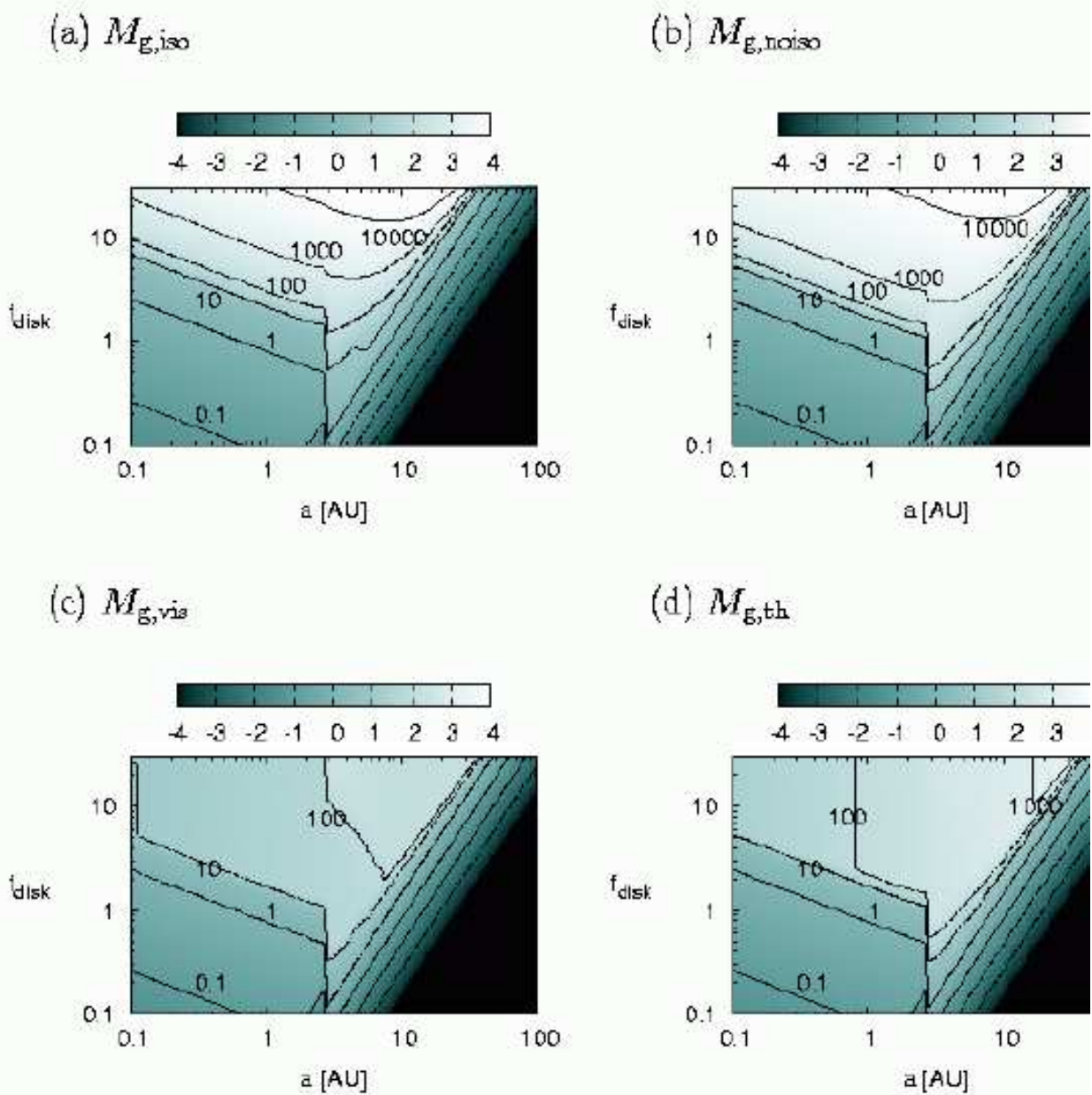}
\caption{
The the same plots as Figures~\ref{fig:core_gas_iso}
except for core truncation by $M_{\rm c,noiso}$.
}
\label{fig:core_gas_noiso}
\end{figure}

\begin{figure}
\plotone{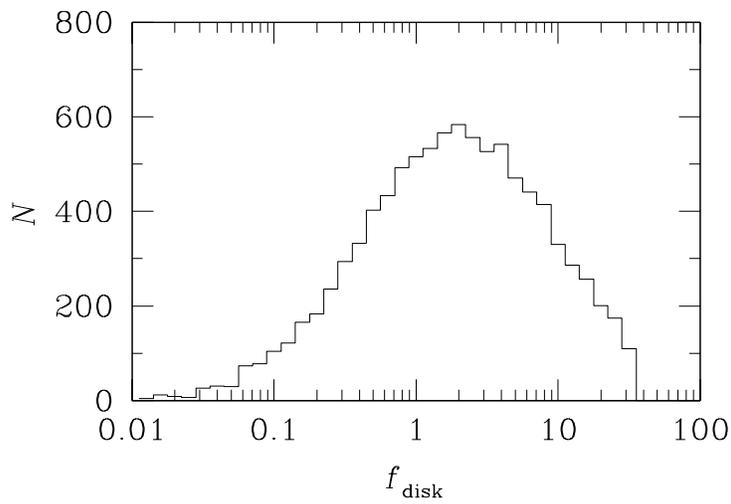}
\caption{
The $f_{\rm disk}$ distribution we use, which
is a gaussian distribution in terms of $\log_{10} f_{\rm disk}$
with a center at $\log_{10} f_{\rm disk} = 0.25$ and dispersion of 1. 
We omit the high $f_{\rm disk}$ tail at $f_{\rm disk} > 30$, since
such heavy disks are self gravitationally unstable. 
}
\label{fig:f_disk_dist}
\end{figure}

\begin{figure}
\plotone{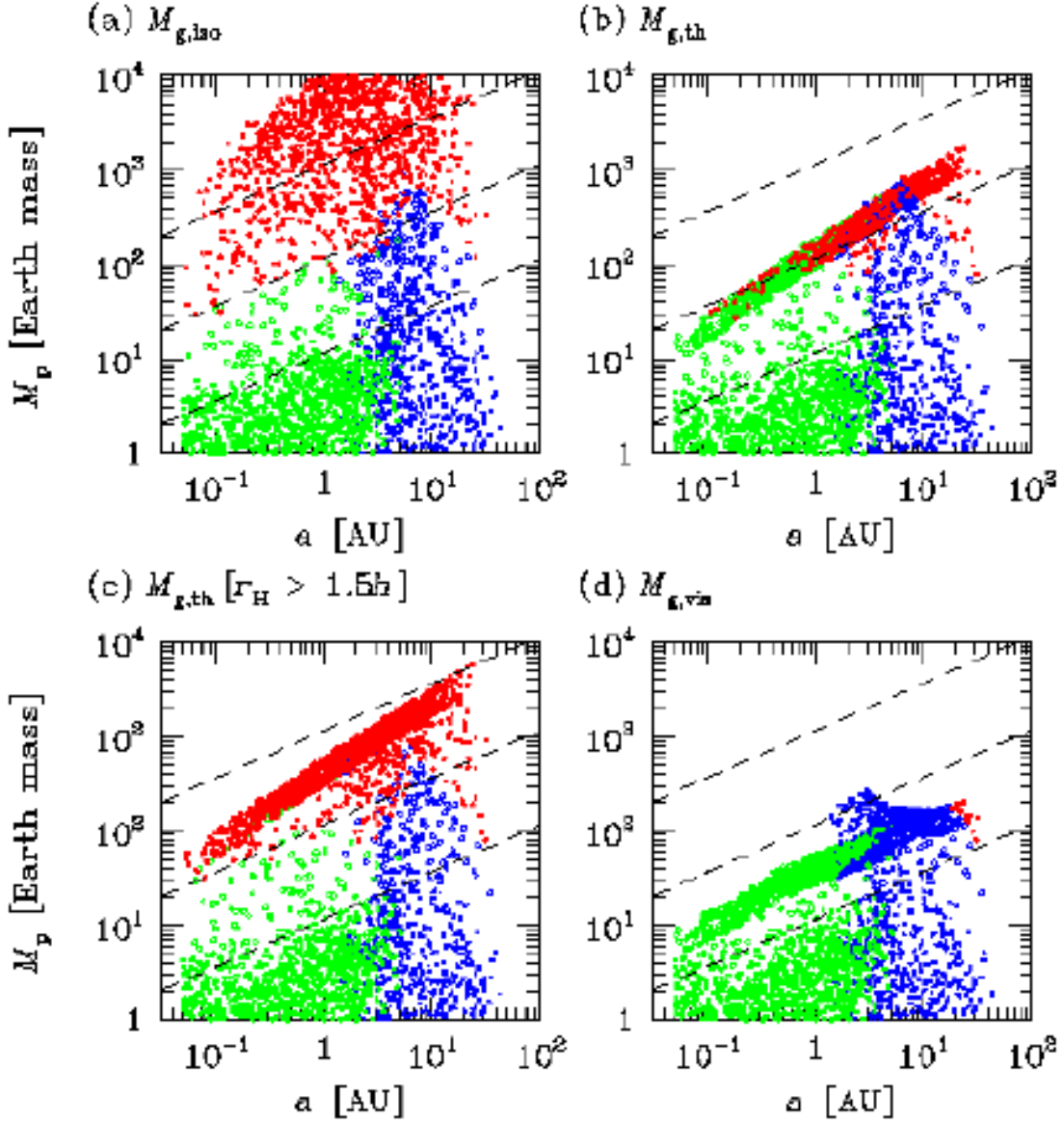}
\caption{
Theoretically predicted distribution based on
core accretion model for gas giant planets 
(for the range of parameters
we used, see text).
Cores are truncated by $M_{\rm c,iso}$.  Gas accretion
is truncated by (a) $M_{\rm g,iso}$, (b) and (c) $M_{\rm g,th}$, and (d)
$M_{\rm g,vis}$.  We adopt $\Delta a_{\rm g} = 2r_{\rm H}$ in (a), the
critical Hill's radius $r_{\rm H,c}$ being $h$ and $1.5h$ in (b) and
(c), respectively, and $\alpha = 10^{-3}$ in (d).
The green filled circles and the blue crosses represent
rocky and icy planets with gaseous
envelopes less massive than their cores.  
The green and blue open circles represent
gas-rich rocky and icy planets with gaseous envelopes which are one to
ten times more massive than their cores.  The red filled circles
represent gas giants with 
envelopes more massive than ten times of their cores.  
For comparison, we also plot observational data of
extrasolar planets in (d).
The dashed ascending lines correspond to radial velocity amplitude of
100 ms$^{-1}$ (upper), 10 ms$^{-1}$ (middle), and 1 ms$^{-1}$ (lower), 
respectively,
assuming the host star mass is $1M_{\odot}$.
}
\label{fig:ma}
\end{figure}

\begin{figure}
\plotone{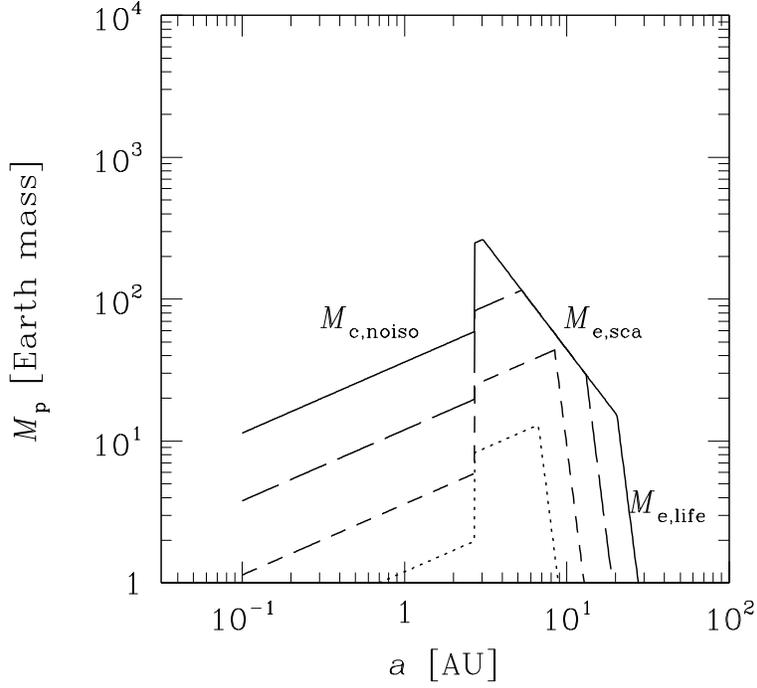}
\caption{
The limits of masses of rocky and icy planets.
The masses are limited by the minimum of $M_{\rm c, noiso}$
(eq.~[\ref{eq:no_m_iso0}]), $M_{\rm e, sca}$ (eq.~[\ref{eq:msca}]),
and $M_{\rm e, life}$ with $t=1$G years (eq.~[\ref{eq:mlife}]).
The solid, long-dashed, short-dashed, and dotted lines
express the cases of $f_{\rm disk} = 30, 10, 3,$ and 1, respectively.
}
\label{fig:ice_giants}
\end{figure}

\begin{figure}
\plotone{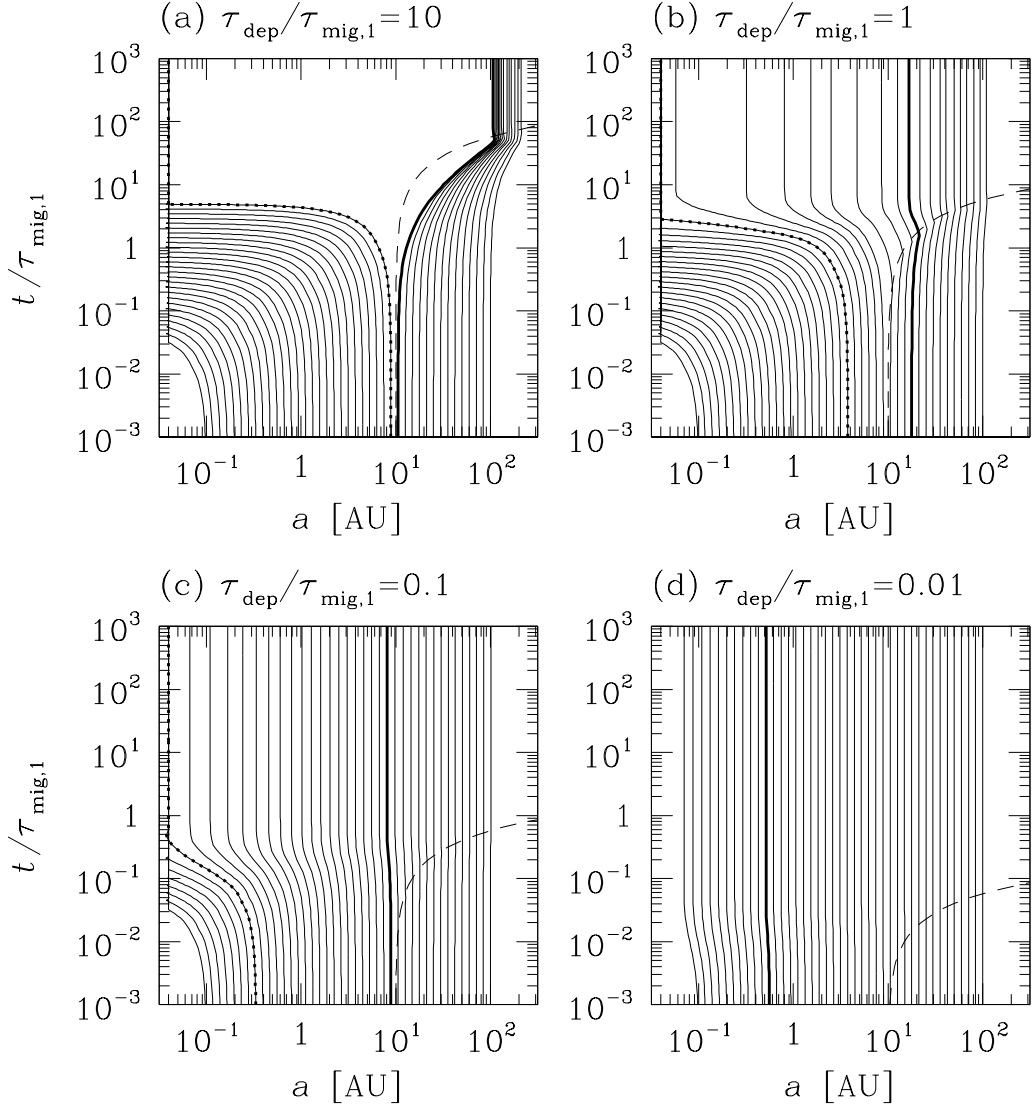}
\caption{
Planetary migration obtained by
integrating $\dot{a}_{\rm p}$
with eqs.~(\ref{eq:tau_mig}) and (\ref{eq:R_m}) is
shown in solid lines. 
Inside the thick dashed lines, final semi major axes reach 0.04AU. 
They are $> 0.9$ times their original locations, outside the 
thick solid lines.
Dashed lines express $R_{\rm m}$.
Time is scaled by
$\tau_{\rm mig}$ at 1AU ($\tau_{\rm mig,1}$).
(a) $\tau_{\rm dep}/\tau_{\rm mig} =10$,
(b) $\tau_{\rm dep}/\tau_{\rm mig} =1$,
(c) $\tau_{\rm dep}/\tau_{\rm mig} =0.1$, and
(d) $\tau_{\rm dep}/\tau_{\rm mig} =0.01$.
}
\label{fig:t_mig}
\end{figure}

\begin{figure}
\plotone{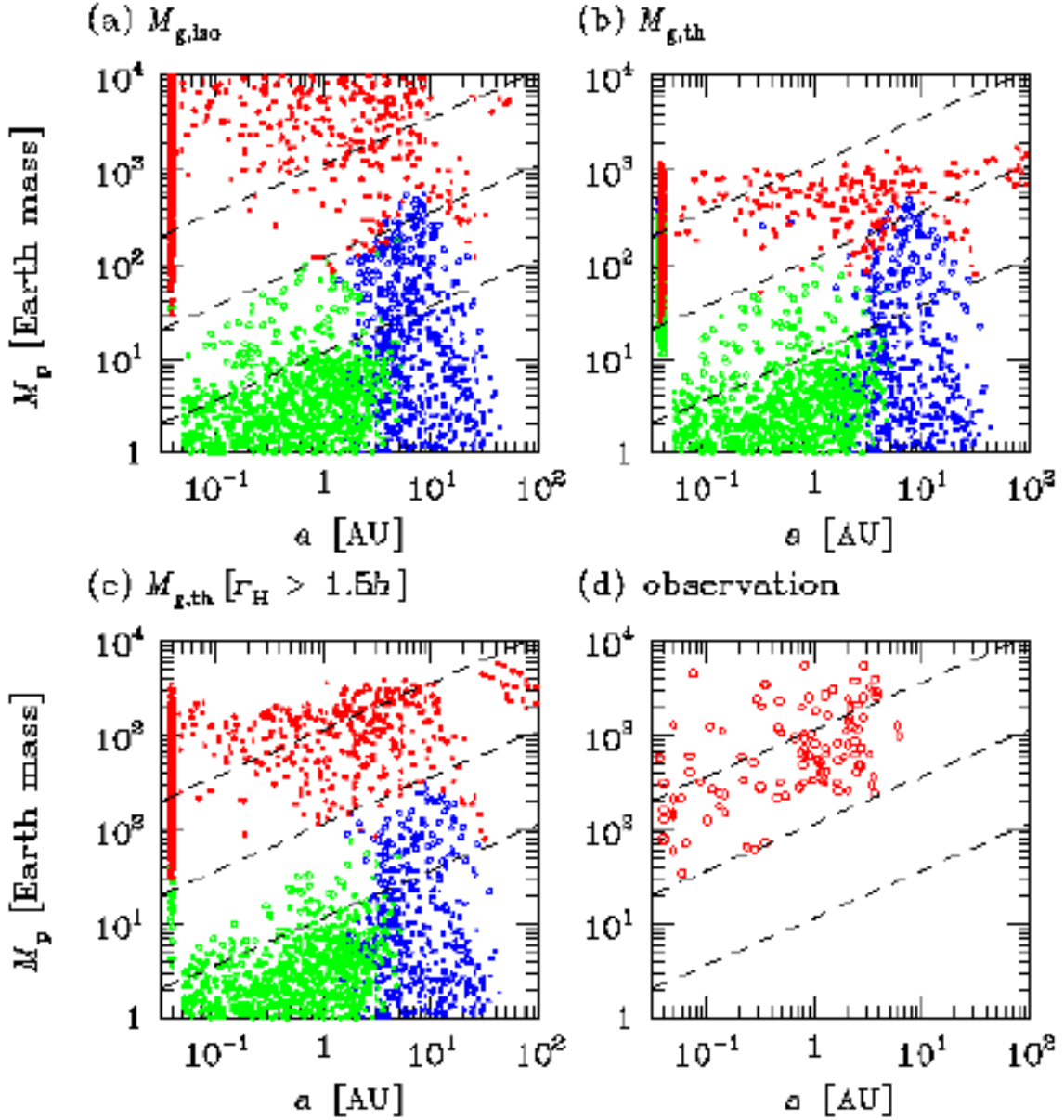}
\caption{
Similar plots as Figures~\ref{fig:ma},
but the effect of type II migration is included.
The value of $\alpha$-viscosity is taken as
$\alpha = 10^{-4}$ to be consistent with disk depletion
times $\sim 10^{6}-10^7$ years.
(a) Gas accretion is truncated by $M_{\rm g,iso}$
and core accretion by $M_{\rm c,iso}$, (b) 
$M_{\rm g,iso}$ and $M_{\rm c,noiso}$
and (c) $M_{\rm g,th}$ and $M_{\rm c,iso}$.
We adopt $\Delta a_{\rm g} = 2r_{\rm H}$ in (a),
$M_* = 1 M_{\odot}$ in (c).
}
\label{fig:ma_mig}
\end{figure}

\begin{figure}
\plotone{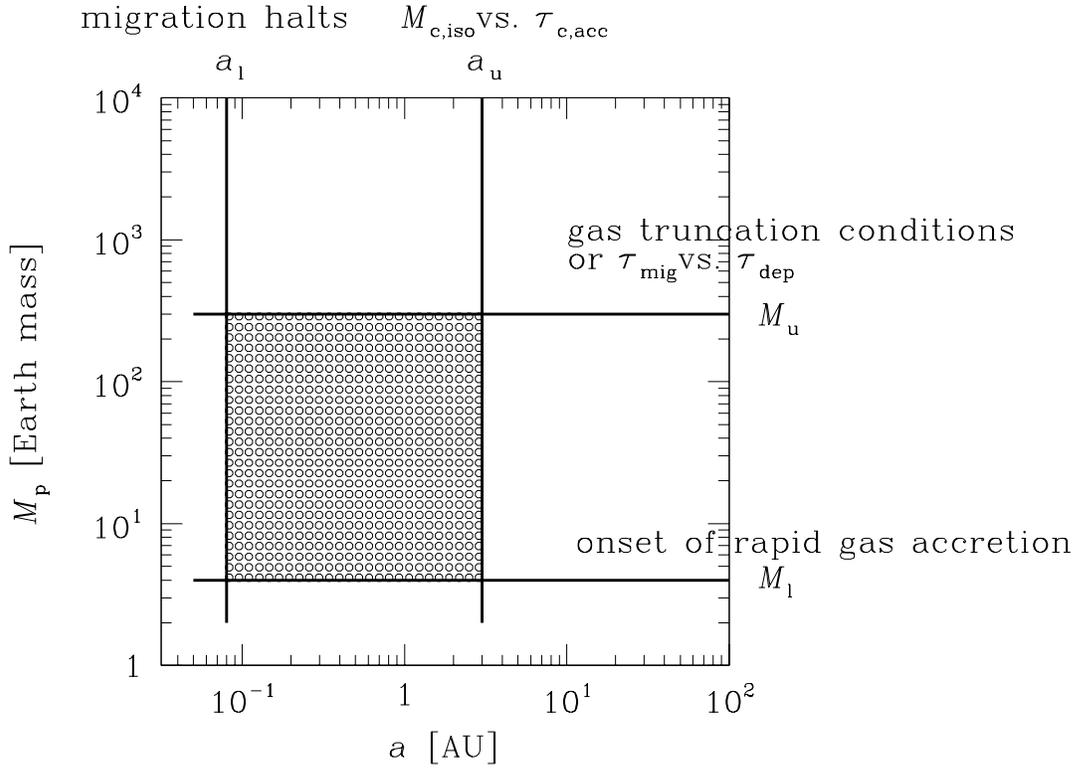}
\caption{
An illustrative figure for the planet desert.
The lower mass boundary ($M_{\rm l}$) indicates the core mass that can
initiate the onset of rapid gas accretion.  
The upper mass boundary ($M_{\rm u}$) and its dependence on
$a$ provide clues on the truncation mechanism of gas accretion.  
The smaller $a$ boundary ($a_{\rm l}$)
indicates the maximum radius at which type II migration can be
halted.
The larger $a$ boundary ($a_{\rm u}$) traces out 
the radius where a dominant condition for 
formation of gas giants changes between $\tau_{\rm c,acc}$
condition and $M_{\rm c,iso}$ condition.
For details, see text.
}
\label{fig:desert}
\end{figure}

\clearpage

\begin{table}

\begin{tabular}{lll} \hline\hline
variables & meaning & definition \\ \hline\hline
$M_{\rm c}$ & a core mass &          \\
$M_{\rm e}$ & an embryo mass & see \S 2.4 \\
$M_{\rm c,iso}$ & a core isolation mass & eq.~(\ref{eq:m_iso}) \\
$M_{\rm c,noiso}$ & a maximum mass of a core & eq.~(\ref{eq:m_non_iso}) \\
$M_{\rm e,iso}$ & an embryo isolation mass & eq.~(\ref{eq:m_e_iso}) \\
$M_{\rm e,sca}$ & an embryo mass limited by scattering & eq.~(\ref{eq:msca}) \\
$M_{\rm e,life}$ & an embryo mass at $t=\tau_*$ (e,g., 1G years) & eq.~(\ref{eq:mlife}) \\
$M_{\rm c,crit}$ & a critical core mass for initiating gas accretion onto the core & eq.~(\ref{eq:crit_core_mass_we_use}) \\
$M_{\rm c,KH}$ & a core mass for K-H contraction time being 
$= \tau_{\rm dep}$ 
& eq.~(\ref{eq:mkh}) \\

\hline
$M_{\rm p}$ & a planet mass including core and envelope &          \\
$M_{\rm g,iso}$ & a gas giant isolation mass & eq.~(\ref{eq:m_gas_iso}) \\
$M_{\rm g,noiso}$ & a maximum mass of a gas giant & eq.~(\ref{eq:m_gas_non_iso}) \\
$M_{\rm g,th}$ & a gas giant truncation mass by thermal condition & eq.~(\ref{eq:m_gas_therm}) \\
$M_{\rm g,vis}$ & a gas giant truncation mass by viscous condition & eq.~(\ref{eq:m_gas_vis}) \\
$M_{\rm g,feed}$ & a supplied gas mass into a feeding zone by viscous diffusion 
 & eq.~(\ref{eq:m_gas_feed}) \\

\hline
$\tau_{\rm c,acc}$ & a core accretion time scale before gas depletion & eq.~(\ref{eq:T_core_grow}) \\
$\tau_{\rm e,acc}$ & an embryo accretion time scale after gas depletion & eq.~(\ref{eq:mdotembr}) \\
$\tau_{\rm dep}$ & a depletion time scale of disk gas & eq.~(\ref{eq:gas_decay}) \\
$\tau_{\rm KH}$ & a Kelvin-Helmholtz contraction time scale of gas envelope & eq.~(\ref{eq:gas_acc_rate_we_use}) \\
$\tau_{\rm mig}$ & a migration time scale of a giant planet & eq.~(\ref{eq:tau_mig}) \\

\hline
$\Sigma_{\rm d}$ & dust surface density of a disk &  \\
$\Sigma_{\rm g}$ & gas surface density of a disk &  \\
$\Sigma_{\rm g,0}$ & initial gas surface density of a disk & eq.~(\ref{eq:gas_decay}) \\
$\eta_{\rm ice}$ & an enhancement factor of $\Sigma_{\rm d}$ due to ice condensation  & eq.~(\ref{eq:sigma_dust})\\
$f_{\rm d}$ & an enhancement factor of $\Sigma_{\rm d}$ from a nominal disk &  
eq.~(\ref{eq:sigma_dust})\\
$f_{\rm g}$ & an enhancement factor of $\Sigma_{\rm g}$ from a nominal disk &  
eq.~(\ref{eq:sigma_g})\\
$f_{\rm g,0}$ & an initial enhancement factor of $\Sigma_{\rm g}$ from a nominal disk 
& eq.~(\ref{eq:fgas_decay}) \\
$f_{\rm disk}$ & $= f_{\rm g,0} = f_{\rm d}$ in this paper &  \\
$f_{\rm tg}$ & a value of $f_{\rm disk}$ to segregate terrestrial/gas-giant planets regions & eq.~(\ref{eq:giants_form_cond1}) or (\ref{eq:giants_form_cond2})  \\
$f_{\rm gi}$ & that to segregate gas/ice giants regions & eq.~(\ref{eq:giants_form_cond3}) \\

\hline
$a_{\rm ice}$ & ice boundary (a semi major axis of ice condensation) &
eq.~(\ref{eq:a_ice}) \\
$a_{\rm tg}$ & a semi major axis to segregate terrestrial$/$gas-giant planets regions & eq.~(\ref{eq:atg}) \\
$a_{\rm gi}$ & that to segregate gas$/$ice giants regions & eq.~(\ref{eq:agi}) \\ \hline
\end{tabular}
\caption{
List of notations
}
\end{table}

\end{document}